\documentclass[12pt,aps,preprint,nofootinbib]{revtex4}

\usepackage{graphicx}
\usepackage{cancel}
\usepackage{amssymb}
\usepackage{textcomp}
\usepackage{amsmath}
\usepackage{bm}
\usepackage{times}
\usepackage{epsfig}
\usepackage{epstopdf}
\usepackage{color}
\usepackage{multirow}
\usepackage[colorlinks=true, pdfstartview=FitV, linkcolor=blue, citecolor=blue, urlcolor=blue]{hyperref}

\usepackage{caption}
\usepackage{subfig}

\def\lsim{\mathrel{\raise.3ex\hbox{$<$\kern-.75em\lower1ex\hbox{$\sim$}}}}
\def\gsim{\mathrel{\raise.3ex\hbox{$>$\kern-.75em\lower1ex\hbox{$\sim$}}}}

\def\beq{\begin{equation}}
\def\eeq{\end{equation}}
\def\be{\begin{equation}}
\def\ee{\end{equation}}
\def\bea{\begin{eqnarray}}
\def\eea{\end{eqnarray}}

\def\to{\rightarrow}

\begin{document}
\title{Singlet Dark Matter in Type II Two Higgs Doublet Model}
\author{Yi Cai$^{\bf a}$}
\email{yi.cai@unimelb.edu.au}

\author{Tong Li$^{\bf b}$}
\email{tong.li@monash.edu}

\affiliation{
$^{\bf a}$ ARC Centre of Excellence for Particle
Physics at the Terascale, School of Physics, University of Melbourne, Melbourne, Victoria 3010, Australia\\
$^{\bf b}$ ARC Centre of Excellence for Particle Physics at the
Terascale, School of Physics, Monash University, Melbourne, Victoria 3800, Australia}

\begin{abstract}
Inspired by the dark matter searches in the low mass region, we
study the Type II two Higgs doublet model with a light gauge singlet
WIMP stabilized by a $Z_2$ symmetry. The real singlet is required to
only couple to the non-Standard Model Higgs. We investigate singlet
candidates with different spins as well as isospin violating effect.
The parameter space favored by LHC data in two Higgs doublet model
and hadronic uncertainties in WIMP-nucleon elastic scattering are
also taken into account. We find only the scalar singlet in the
isospin conserving case leads to a major overlap with the region of
interests of most direct detection experiments.
\end{abstract}

\maketitle
\section{Introduction}
\label{sec:intro}

The existence of some mysterious component of the Universe, namely the dark matter (DM),
has been solidified by multiple astrophysical and cosmological observations.
The weakly interacting massive particle (WIMP) is one popular candidate of DM.
Recently various underground direct detection experiments such as
DAMA~\cite{DAMA}, CoGeNT~\cite{COGENT}, CRESST~\cite{CRESST} and
CDMS~\cite{CDMS} have shown some DM-like events in the low mass region.
Although there is still no conclusive statement for the existence of WIMP based on these events,
it is worth paying special attention to the light WIMP considering undergoing searches in low mass region.


Since there is no viable WIMP candidates in the Standard Model (SM),
extensions of this highly successful theory is necessary. One simple
extension of the SM is to add a real SM gauge singlet with a mass
of the electroweak scale or less~\cite{singlet1,singlet2,singlet3}.
An unbroken $Z_2$ parity, under which only the singlet is odd, stabilizes this particle (called $D$ here).

At the renormalizable level the singlet only couples to the SM Higgs doublet.
The coupling has to be carefully adjusted to reproduce the required relic density of $D$,
while satisfying the constraints from the indirect and direct searches~\cite{SMsinglet,darkon,hedarkon}.
It is, however, a hard task to accomplish within the SM+D framework,
taking into account the results from the direction detection of DM and the search for invisible decay
mode of SM Higgs at the LHC. For recent discussions, see Ref.~\cite{update} and references therein.
In order to achieve a consistent
scenario with singlet dark matter, it is desirable to consider an
extension of the SM, such as the Two Higgs doublet model (THDM) that
we discuss here~\cite{hedarkon,2hdmd}.


In this paper we study the THDM+D framework, with singlet mass
$\lesssim 20$ GeV, in light of the implications of the LHC Higgs
search results for the Type II THDM. To accommodate the SM Higgs search
data, the coupling between the singlet and the SM-like Higgs is forbidden by hand
and DM $D$ can only be produced in pairs via the non-SM Higgs boson.
In particular the isospin-violating effect introduced to reconcile
the experiments above and XENON100 is discussed~\cite{jfeng}, in terms of the
allowed parameter region in Type II THDM. We investigate three
candidates of real singlet field, namely scalar, Majorana fermion
and vector.

The paper is organized as follows. In Sec.~\ref{sec:singletTHDM} we describe the
properties of THDM+D framework, where we also discuss the constraints on Type II THDM from the LHC Higgs search.
The results of $D$-Higgs scattering
interaction are presented in Sec.~\ref{sec:results}. In this section we also display the $D$-nucleon elastic cross sections. We summarize our conclusions in Sec.~\ref{sec:conclusion}.

\section{Singlet Dark Matter in Type II THDM}
\label{sec:singletTHDM}

\subsection{Model and Experimental Constraints}
The interactions between the scalar $S$, Majorana fermion $\chi$, vector
$V_\mu$ singlet and Higgs sector in the Type II THDM are
\begin{eqnarray}
\mathcal{L}_S&=&-{1\over 2}S^2(\lambda_{S1}H_1^\dagger H_1+\lambda_{S2}H_2^\dagger H_2)-{m_{S0}^2\over 2}S^2-{\lambda_S'\over 4!}S^4,\\
\mathcal{L}_{\chi}&=&-{1\over 2\Lambda}\overline{\chi^c}\chi (\lambda_{\chi 1}H_1^\dagger H_1
+\lambda_{\chi 2}H_2^\dagger H_2)-{m_\chi\over 2}\overline{\chi^c}\chi+h.c.,\\
\mathcal{L}_V&=&{1\over 2}V_\mu V^\mu (\lambda_{V1}H_1^\dagger
H_1+\lambda_{V2}H_2^\dagger H_2)+{m_V^2\over 2}V_\mu
V^\mu-{\lambda_V'\over 4!}(V_\mu V^\mu)^2-{1\over
4}V^{\mu\nu}V_{\mu\nu},
\end{eqnarray}
where $\Lambda$ is a dimensional scale for fermionic singlet. Note that
a discrete $Z_2'$ symmetry, under which only $H_2$ is odd, is
introduced here to forbid other Higgs interactions. The two Higgs
doublets are decomposed as
\begin{eqnarray}
&&H_i=\left(
      \begin{array}{c}
        h_i^+ \\
        (v_i+h_i+iP_i)/\sqrt{2} \\
      \end{array}
    \right) \ i=1,2, \quad  \left(
      \begin{array}{c}
        h_1^+ \\
        h_2^+ \\
      \end{array}
    \right)
= \left(
  \begin{array}{cc}
    \cos\beta & -\sin\beta \\
    \sin\beta & \cos\beta \\
  \end{array}
\right) \left(
      \begin{array}{c}
        G^+ \\
        H^+ \\
      \end{array}
    \right), \\
&&\left(
      \begin{array}{c}
        P_1 \\
        P_2 \\
      \end{array}
    \right)
= \left(
  \begin{array}{cc}
    \cos\beta & -\sin\beta \\
    \sin\beta & \cos\beta \\
  \end{array}
\right) \left(
      \begin{array}{c}
        G^0 \\
        A^0 \\
      \end{array}
    \right), \quad
\left(
      \begin{array}{c}
        h_1 \\
        h_2 \\
      \end{array}
    \right)
= \left(
  \begin{array}{cc}
    \cos\alpha & -\sin\alpha \\
    \sin\alpha & \cos\alpha \\
  \end{array}
\right) \left(
      \begin{array}{c}
        H^0 \\
        h^0 \\
      \end{array}
    \right),
\end{eqnarray}
with $\tan\beta=v_2/v_1$, $v_0^2=v_1^2+v_2^2\approx (246 \ {\rm
GeV})^2$ and $\alpha$ being the mixing angle between two neutral
CP-even Higgses. After the electroweak symmetry breaking, one gets
the following DM interactions with the neutral CP-even Higgses
\begin{eqnarray}
\mathcal{L}_{SSh}&=& -(-\lambda_{S1}\sin\alpha\cos\beta+\lambda_{S2}\cos\alpha\sin\beta)v_0S^2h^0/2\equiv-\lambda_{Sh} v_0S^2h^0/2,\\
\mathcal{L}_{SSH}&=&
-(\lambda_{S1}\cos\alpha\cos\beta+\lambda_{S2}\sin\alpha\sin\beta)v_0S^2H^0/2\equiv-\lambda_{SH}
v_0S^2H^0/2,
\end{eqnarray}
for scalar candidate $S$,
\begin{eqnarray}
\mathcal{L}_{FFh}&=& -(-\lambda_{\chi 1}\sin\alpha\cos\beta+\lambda_{\chi 2}\cos\alpha\sin\beta)v_0\overline{\chi^c}\chi h^0/\Lambda\equiv\lambda_{Fh} F^2h^0/2,\\
\mathcal{L}_{FFH}&=& -(\lambda_{\chi
1}\cos\alpha\cos\beta+\lambda_{\chi
2}\sin\alpha\sin\beta)v_0\overline{\chi^c}\chi
H^0/\Lambda\equiv-\lambda_{FH} F^2H^0/2,
\end{eqnarray}
for fermion candidate $F=\chi+\chi^c$ and
\begin{eqnarray}
\mathcal{L}_{VVh}&=& (-\lambda_{V1}\sin\alpha\cos\beta+\lambda_{V2}\cos\alpha\sin\beta)v_0V^2h^0/2\equiv\lambda_{Vh} v_0V^2h^0/2,\\
\mathcal{L}_{VVH}&=&
(\lambda_{V1}\cos\alpha\cos\beta+\lambda_{V2}\sin\alpha\sin\beta)v_0V^2H^0/2\equiv\lambda_{VH}
v_0V^2H^0/2,
\end{eqnarray}
for vector candidate $V$ respectively. The size of the couplings
$\lambda_{D\mathcal{H}} \ (D=S,F,V; \mathcal{H}=h^0,H^0)$ controls
the signal strength of the DM detection and is determined by the constraints
from DM relic density measurement and the implication of SM Higgs searches for THDM at LHC.

In the Type II THDM, the interactions between two CP-even Higgses
and massive gauge bosons follow the same behavior as in the Minimal
Supersymmetric Standard Model (MSSM):
\begin{eqnarray}
\mathcal{L}_{\mathcal{H}WW/ZZ}&=&\left({2m_W^2\over
v_0}WW+{m_Z^2\over
v_0}ZZ\right)\left(h^0\sin(\beta-\alpha)+H^0\cos(\beta-\alpha)\right),
\end{eqnarray}
where SM-like Higgs is the one which couples to the gauge bosons more
strongly. Current studies of the LHC data and the Type II THDM imply the
existence of two scenarios~\cite{2hdmcons}
\begin{eqnarray}
\nonumber
I:&& \sin(\beta-\alpha)\approx \pm 1, h^0 \ {\rm SM-like},\\
&& 100 <M_{H^0}<750
\ {\rm GeV}, 300 <M_{H^\pm}<800
\ {\rm GeV}, M_{A}<800 \ {\rm GeV};  \\
\nonumber
II:&& \sin(\beta-\alpha)\approx 0, H^0 \ {\rm SM-like},\\
&&70<M_{h^0}<126 \ {\rm GeV}, M_A\simeq M_{H^+}>300 \ {\rm GeV},
\end{eqnarray}
with $0.5<\tan\beta<4$ in both cases. Unlike the MSSM, in Type II THDM,
the masses of $H^0, A^0, H^\pm$ are largely uncorrelated. In order
to simplify the non-SM Higgs decay modes for concreteness, we
take the degenerate case, namely $M_{H^0}\approx M_{A^0}\approx M_{H^\pm}$, in
scenario I.


\subsection{Singlet DM Annihilation and Singlet-Nucleon Scattering Interaction}

As mentioned in the introduction, in order to accommodate a small SM
Higgs invisible decay branching ratio, we set vanishing couplings
$\lambda_{Dh^0} (\lambda_{DH^0})=0$ in scenario I (II). Thus the DM
singlets can only be produced in pairs from non-SM Higgs decay. The
partial widths of the Higgs boson $\mathcal{H}$ decay into DM
singlet are given by~\cite{various}
\begin{eqnarray}
\Gamma_S&=&{\lambda_{S\mathcal{H}}^2v_0^2\over 32\pi m_\mathcal{H}}\sqrt{1-{4m_S^2\over m_\mathcal{H}^2}},\\
\Gamma_F&=&{\lambda_{F\mathcal{H}}^2m_\mathcal{H}\over 16\pi}\left(1-{4m_F^2\over m_\mathcal{H}^2}\right)^{3/2},\\
\Gamma_V&=&{\lambda_{V\mathcal{H}}^2v_0^2m_\mathcal{H}^3\over 128\pi
m_V^4}\left(1-{4m_V^2\over m_\mathcal{H}^2}+{12m_V^4\over
m_\mathcal{H}^4}\right)\sqrt{1-{4m_V^2\over m_\mathcal{H}^2}}.\label{eqn:gammav}
\end{eqnarray}
The $\mathcal{H}$-mediated annihilation cross sections of DM
singlets are thus
\begin{eqnarray}
\label{eqn:anns}
\sigma_{ann} v_{rel}(S)&=&{2\lambda_{S\mathcal{H}}^2v_0^2\over
(4m_S^2-m_\mathcal{H}^2)^2+\Gamma_\mathcal{H}^2m_\mathcal{H}^2}
{\sum_i\Gamma(\tilde{\mathcal{H}}\to X_i)\over 2m_S},\\
\label{eqn:annf}
\sigma_{ann} v_{rel}(F)&=&{\lambda_{F\mathcal{H}}^2m_F^2\over
(4m_F^2-m_\mathcal{H}^2)^2+\Gamma_\mathcal{H}^2m_\mathcal{H}^2}
{\sum_i\Gamma(\tilde{\mathcal{H}}\to X_i)\over 2m_F}v_{rel}^2,\\
\label{eqn:annv}
\sigma_{ann} v_{rel}(V)&=&{2\lambda_{V\mathcal{H}}^2v_0^2/3\over
(4m_V^2-m_\mathcal{H}^2)^2+\Gamma_\mathcal{H}^2m_\mathcal{H}^2}
{\sum_i\Gamma(\tilde{\mathcal{H}}\to X_i)\over 2m_V},
\end{eqnarray}
where $v_{rel}=2|p^{cm}_D|/m_D$ is the relative speed of the DM pair
in their center-of-mass frame, $\tilde{\mathcal{H}}$ is a virtual
Higgs boson with the same couplings to other particles as the
physical $\mathcal{H}$ of mass $m_\mathcal{H}$, but with an
invariant mass $\sqrt{s}=2m_D$, and $\tilde{\mathcal{H}}\to X_i$ is
any possible decay mode of $\mathcal{H}$ except that into dark matter.

As the non-SM Higgs does not couple to gauge bosons, it can only decay into fermion pairs and loop processes
induced by Yukawa couplings besides DM
singlet. In the Type II THDM, the Yukawa
lagrangian is given by
\begin{eqnarray}
\mathcal{L}_Y&=&-Y_2^u\overline{Q}_L\tilde{H}_2U_R-Y_1^d\overline{Q}_LH_1D_R-Y_1^l\overline{l}_LH_1E_R+h.c.
\end{eqnarray}
where $U_R$ is also an odd field under $Z_2'$ symmetry. It leads to
the following Yukawa interactions
\begin{eqnarray}
\mathcal{L}_{ff\mathcal{H}}&=&-\bar{U}_LM^u
U_R\left({\cos\alpha\over \sin\beta}{h^0\over v_0}+{\sin\alpha\over
\sin\beta}{H^0\over v_0}\right)-
\bar{D}_LM^d D_R\left(-{\sin\alpha\over \cos\beta}{h^0\over v_0}+{\cos\alpha\over \cos\beta}{H^0\over v_0}\right)\nonumber \\
&&- \bar{E}_LM^l E_R\left(-{\sin\alpha\over \cos\beta}{h^0\over
v_0}+{\cos\alpha\over \cos\beta}{H^0\over v_0}\right)+h.c..
\label{Yukawa}
\end{eqnarray}

These Yukawa couplings also determine the singlet DM-nucleon
scattering process with non-SM Higgs mediation in t-channel.
Following Eq.~(\ref{Yukawa}), the non-SM Higgs
$\mathcal{H}$-$q$-$\bar{q}$ interactions $\kappa_q^\mathcal{H}$ read
\begin{eqnarray}
\kappa_u^H&=&\kappa_c^H=\kappa_t^H={\sin\alpha\over \sin\beta}, \
\kappa_d^H=\kappa_s^H=\kappa_b^H={\cos\alpha\over \cos\beta}, \ \ \ {\rm for \ scenario \ I},
\label{kappai}\\
\kappa_u^h&=&\kappa_c^h=\kappa_t^h={\cos\alpha\over \sin\beta}, \
\kappa_d^h=\kappa_s^h=\kappa_b^h=-{\sin\alpha\over \cos\beta}, \ \ \
{\rm for \ scenario \ II}.
\label{kappaii}
\end{eqnarray}
It turns out that such THDM Yukawa couplings in the two limits are
simplified into those in the MSSM
\begin{eqnarray}
\hspace{-1cm}&&{\sin\alpha\over \sin\beta}=\mp {1\over \tan\beta}, \
{\cos\alpha\over \cos\beta}=\pm \tan\beta, \ \tan\alpha={-1\over
\tan\beta}, \ {\rm for \ \sin(\beta-\alpha)=\pm 1 \ in \ scenario \
I},
\label{I}\\
\hspace{-1cm}&&{\cos\alpha\over \sin\beta}={1\over \tan\beta}, \
-{\sin\alpha\over \cos\beta}=-\tan\beta, \ \tan\alpha=\tan\beta, \
{\rm for \ \sin(\beta-\alpha)=0 \ in \ scenario \ II}. \label{II}
\end{eqnarray}
From Eqs.~(\ref{I}) and (\ref{II}), one can see that the
$\mathcal{H}$-$q$-$\bar{q}$ interactions are independent of the choice of
Higgs scenarios.

Combining the interactions of $\mathcal{H}$-$D$-$D$ and
$\mathcal{H}$-$q$-$\bar{q}$, the elastic cross sections of singlet DM
with proton can be written as
\begin{eqnarray}
\sigma_{el}^p(S)&=&{\lambda_{S\mathcal{H}}^2m_p^2g_{pp\mathcal{H}}^2v_0^2\over 4\pi m_{\mathcal{H}}^4(m_S+m_p)^2}={4m_p^2m_S^2f_{Sp}^2\over \pi (m_S+m_p)^2}, \ \ \ f_{Sp}={\lambda_{S\mathcal{H}}g_{pp\mathcal{H}}v_0\over 4m_S m_\mathcal{H}^2}, \label{elS}\\
\sigma_{el}^p(F)&=&{\lambda_{F\mathcal{H}}^2m_p^2m_F^2g_{pp\mathcal{H}}^2\over \pi m_{\mathcal{H}}^4 (m_F+m_p)^2}={4m_p^2m_F^2f_{Fp}^2\over \pi (m_F+m_p)^2}, \ \ \ f_{Fp}={\lambda_{F\mathcal{H}}g_{pp\mathcal{H}}\over 2m_\mathcal{H}^2},\label{elF}\\
\sigma_{el}^p(V)&=&{\lambda_{V\mathcal{H}}^2m_p^2g_{pp\mathcal{H}}^2v_0^2\over
4\pi m_\mathcal{H}^4(m_V+m_p)^2}={4m_p^2m_V^2f_{Vp}^2\over \pi
(m_V+m_p)^2}, \ \ \
f_{Vp}={\lambda_{V\mathcal{H}}g_{pp\mathcal{H}}v_0\over 4m_V
m_\mathcal{H}^2}, \label{elV}
\end{eqnarray}
where
$g_{pp\mathcal{H}}=\sum_{u,d,s,c,b,t}g_q^p\kappa_q^\mathcal{H}$ and
$m_p$ denotes the mass of proton. If isospin-violating effect is taken into account, one has relation
$f_{Dn}/f_{Dp}=g_{nn\mathcal{H}}/g_{pp\mathcal{H}}$. The relevant
variables in Eqs.~(\ref{elS}), (\ref{elF}) and (\ref{elV}) are
collected in Appendix~\ref{app:gppH} for both isospin-conserving (IC) and
isospin-violating (IV) cases.

\section{Numerical Results}
\label{sec:results}

\begin{figure}[tb]
\begin{center}
\includegraphics[scale=1,width=8cm]{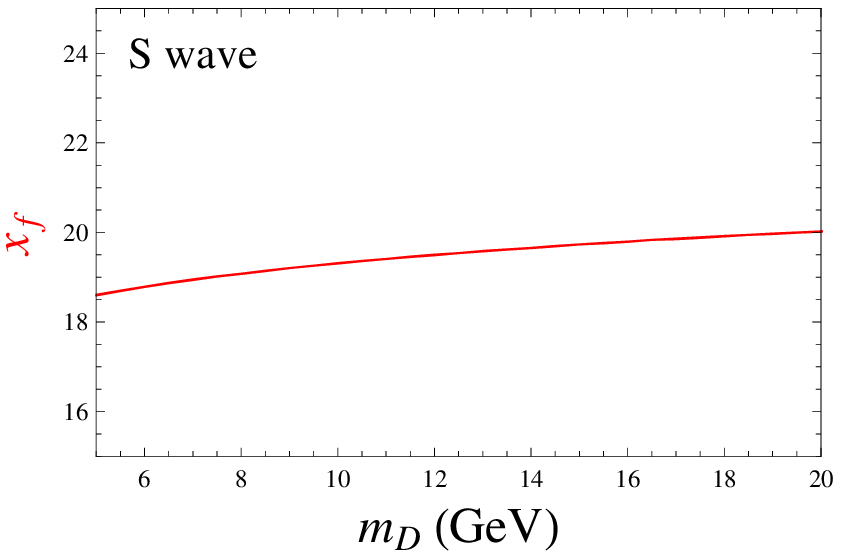}
\includegraphics[scale=1,width=8cm]{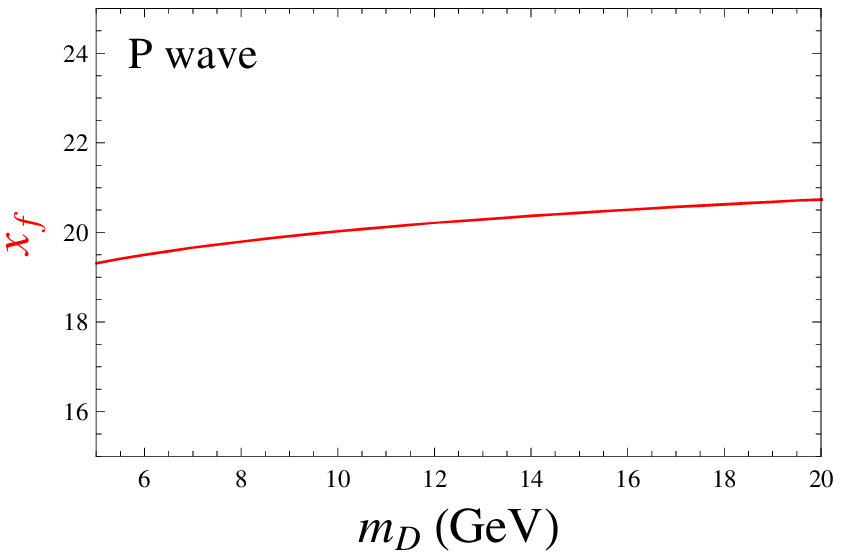}\\
\includegraphics[scale=1,width=8cm]{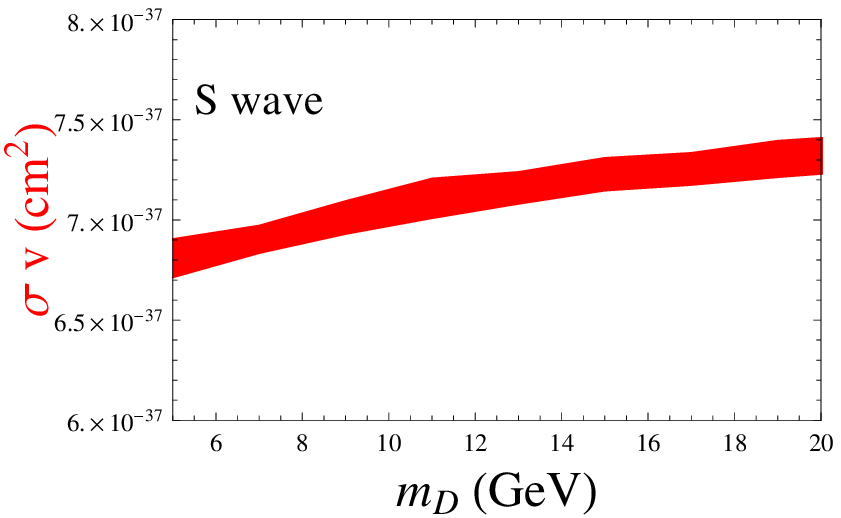}
\includegraphics[scale=1,width=8cm]{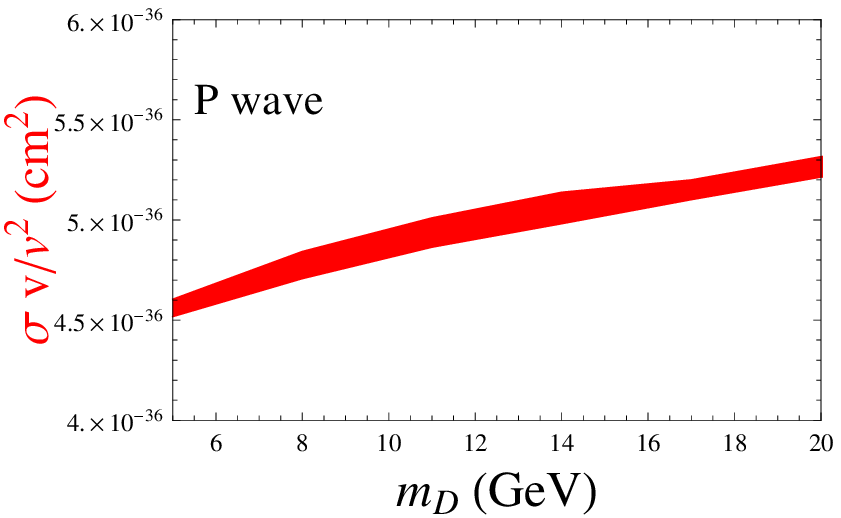}
\end{center}
\caption{$x_f$ and $\sigma_{ann} v_{rel}$ vs. $m_D$ for annihilation in pure S- and P- waves.
} \label{xf}
\end{figure}

For a given interaction between the WIMP and SM particles,
the relic density $\Omega_D$ can be calculated by~\cite{turner}
\begin{eqnarray}
&&\Omega_Dh^2\simeq{1.07\times 10^9 x_f\over
M_{Pl}\sqrt{g_\ast}(a+3b/x_f){\rm GeV}}, \ \ \ x_f\simeq {\rm
ln}{0.038M_{Pl}m_D(a+6b/x_f)\over \sqrt{g_\ast x_f}} \label{omega},
\end{eqnarray}
where $a$ and $b$ are the coefficients of the Taylor expansion in $v_{rel}^2$ from the annihilation cross section $\sigma_{ann} v_{rel}=a+b v_{rel}^2$. Here $h$ is the Hubble constant in the unit of 100 ${\rm km}/(s\cdot Mpc)$,
$M_{Pl}$ is the Plank scale, $x_f=m_D/T_f$ with $T_f$ being the
freezing temperature, and $g_\ast$ is the total number of relativistic degrees of freedom at $T_f$.
In Fig.~\ref{xf} we show the restricted regions of $x_f$ and $\sigma_{ann} v_{rel}$ for pure S- and P-wave annihilation as a function of
the WIMP mass $m_D$. They are obtained from Eq.~(\ref{omega}) with $\Omega_D h^2=0.1187\pm 0.0017$~\cite{planck} and
thus independent of the explicit form of the SM-WIMP interaction as well as other properties of the dark matter candidate. The freeze-out temperatures in both cases are almost the same with the one in S-wave slightly higher.
The coefficient $b$ in P-wave annihilation is nearly one order of magnitude larger than $a$ in the S-wave case, as a result of compensating for the P-wave suppression.

\subsection{Higgs-proton coupling $g_{pp\mathcal{H}}$ in both IC and IV cases}
\label{gppH}
\begin{figure}[tb]
\begin{center}
\includegraphics[scale=1,width=8cm]{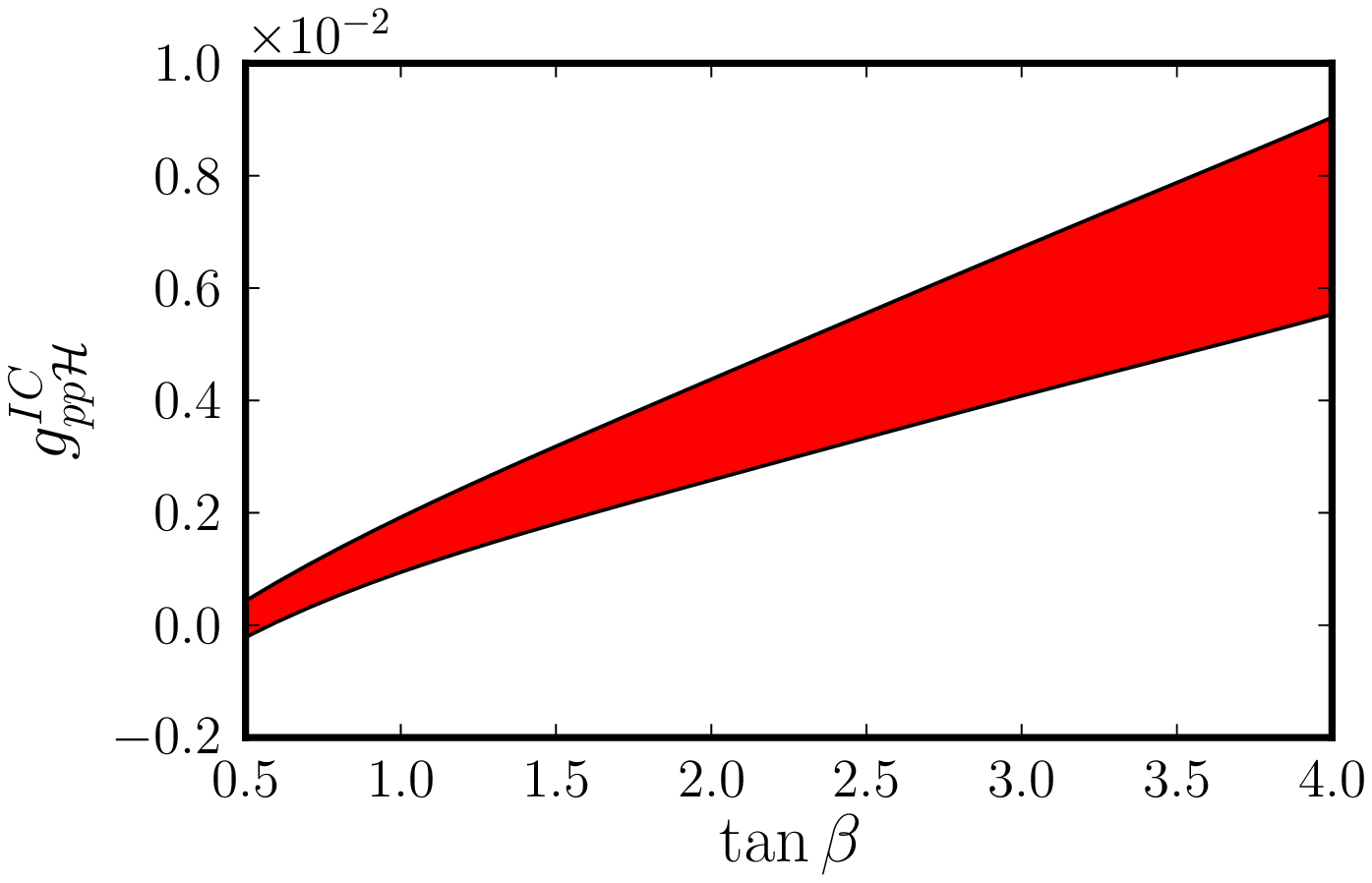}\\
\includegraphics[scale=1,width=8cm]{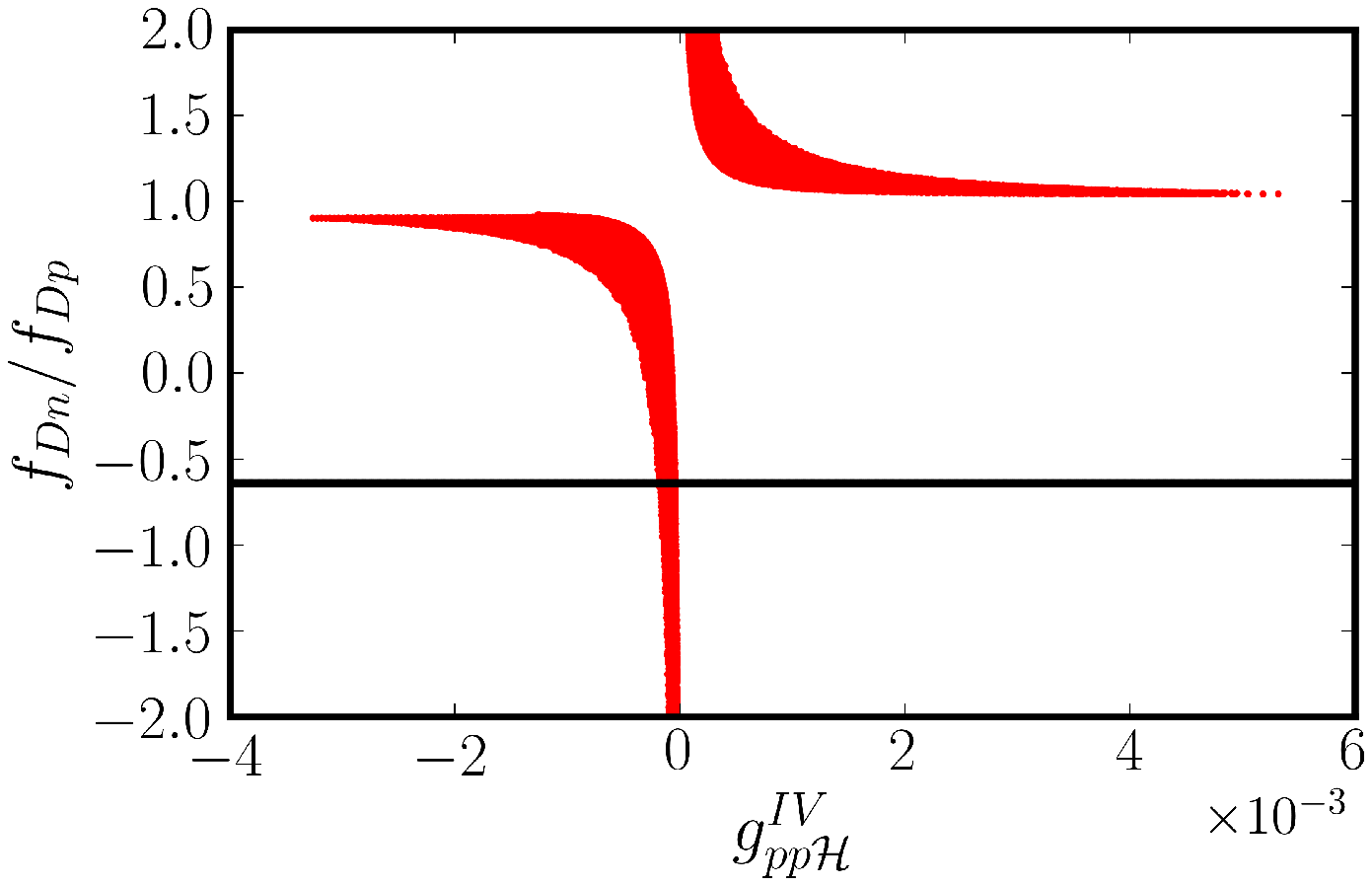}
\includegraphics[scale=1,width=8cm]{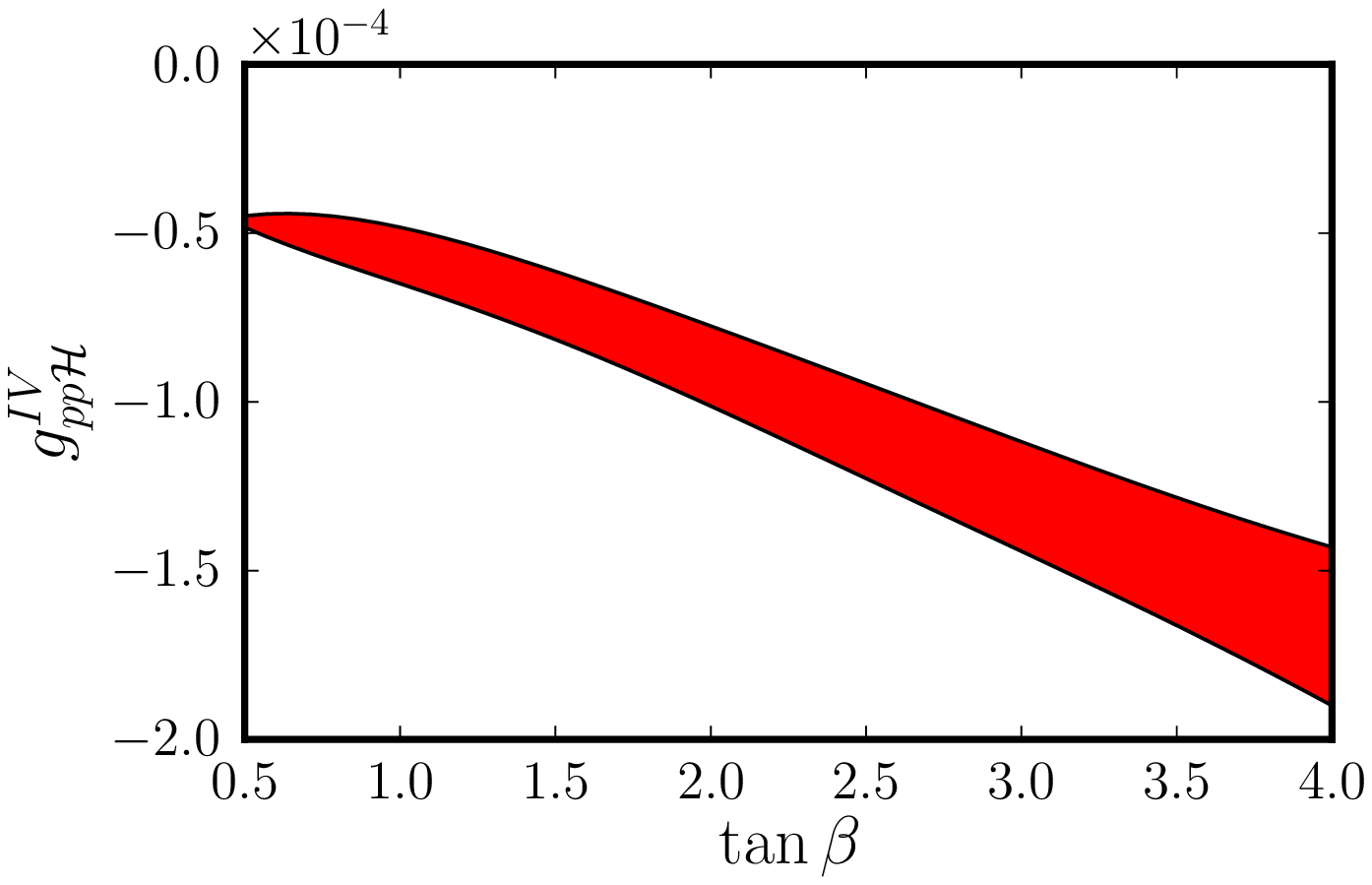}
\end{center}
\caption{Top: $g_{pp\mathcal{H}}$ vs. $\tan\beta$ for IC case
($f_{Dn}/f_{Dp}=1$). Bottom left: the scatter plot of
$f_{Dn}/f_{Dp}$ vs. $g_{pp\mathcal{H}}$ for IV case
($f_{Dn}/f_{Dp}\neq 1$) with $0.5<\tan\beta<4$. The horizontal line
denotes the nearly xenon-phobic value $f_{Dn}/f_{Dp}=-0.64$. Bottom
right: $g_{pp\mathcal{H}}$ vs. $\tan\beta$ for the xenon-phobic
value, $f_{Dn}/f_{Dp}=-0.64$. These values of $g_{pp\mathcal{H}}$
are for $\sin(\beta-\alpha)=-1 \ {\rm or} \ 0$. The sign of the
coupling should be flipped for $\sin(\beta-\alpha)=1$ case.}
\label{fig:gppH}
\end{figure}

Due to the uncertainties of quark masses and hadronic matrix elements, the coupling $g_{pp\mathcal{H}}$
may vary quite a bit.
The dependence of $g_{pp\mathcal{H}}$ on different hadronic quantities is shown in Appendix~\ref{app:gppH}. It is also determined by the Higgs-quark interactions $\kappa_q^\mathcal{H}$, namely $\tan\beta$, as shown in Eqs.~(\ref{kappai}) and (\ref{kappaii}). Thus, its absolute value is independent of different spectrum limits in Higgs sector.
In Fig.~\ref{fig:gppH} we show the coupling $g_{pp\mathcal{H}}$ as a function of $\tan\beta$ in both IC (top) and IV (bottom) cases. One can see that, in the IV case with $f_{Dn}/f_{Dp}=-0.64$, the coupling $g_{pp\mathcal{H}}$ is nearly two
orders of magnitude smaller compared to the one in IC case because of the medium $\tan\beta$ values.

\subsection{$\mathcal{H}$-$D$-$D$ coupling $\lambda_{D\mathcal{H}}$}
\label{subsec:lambdadh}
\begin{figure}[tb]
\begin{center}
\includegraphics[scale=1,width=8cm]{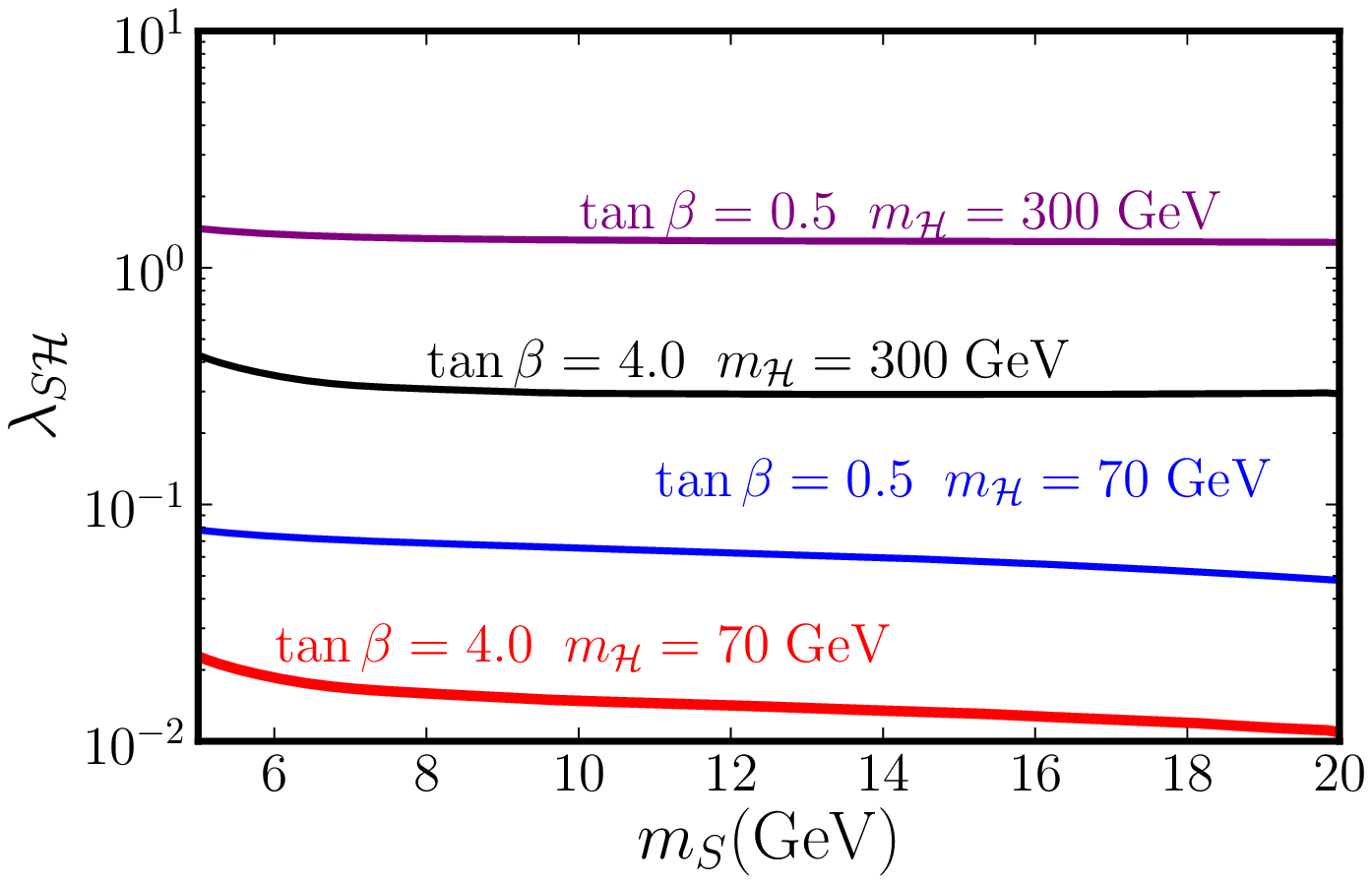}\\
\includegraphics[scale=1,width=8cm]{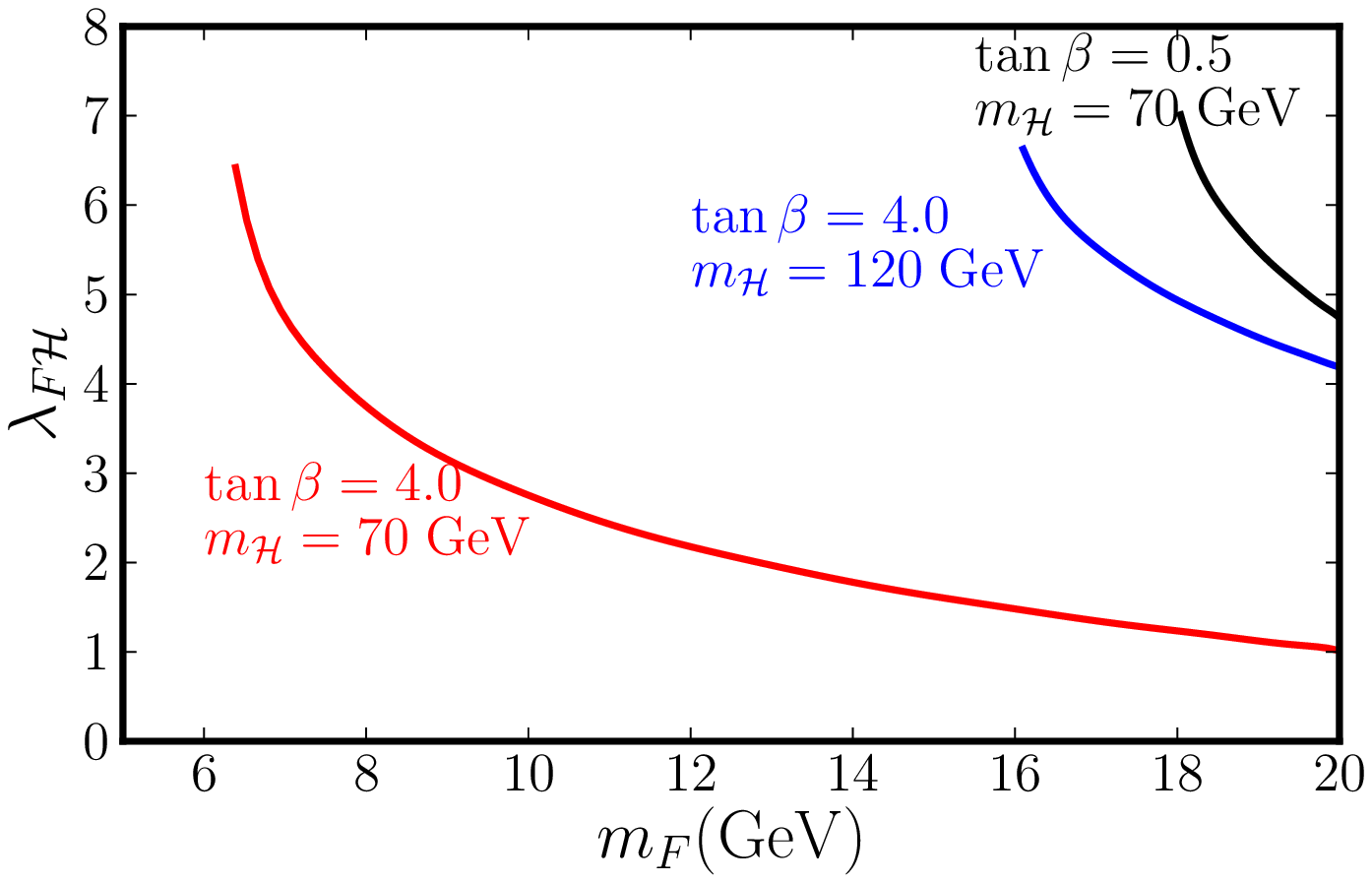}
\includegraphics[scale=1,width=8cm]{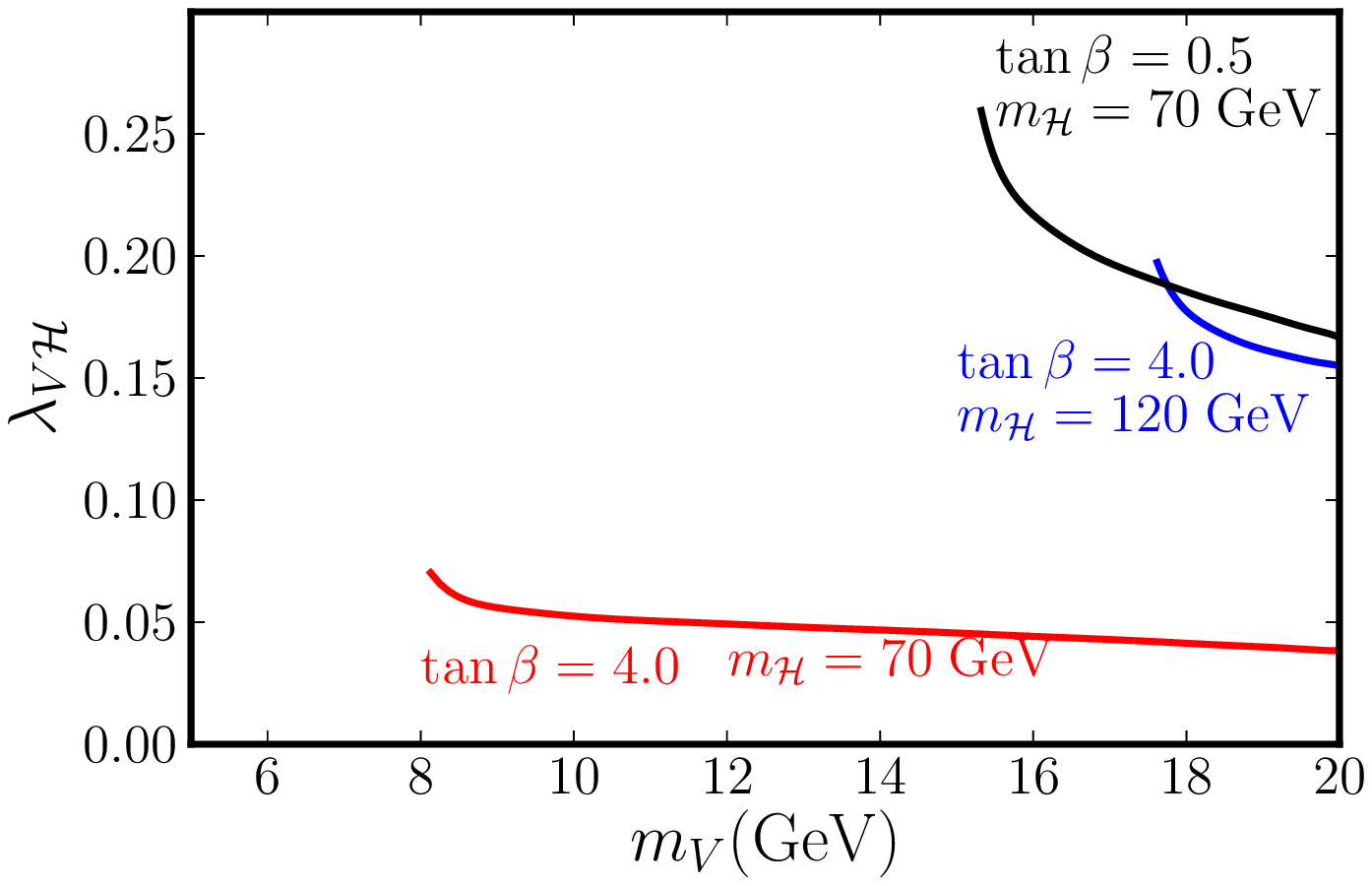}
\end{center}
\caption{Top: $\lambda_{S\mathcal{H}}$ vs. $m_S$ for $\tan\beta=0.5$
and $\tan\beta=4.0$ with $m_{\mathcal{H}}=70\; \rm{GeV}$ and
$m_{\mathcal{H}}=300 \;\rm{GeV}$. Bottom: $\lambda_{F\mathcal{H}}$
vs. $m_F$ (left) and $\lambda_{V\mathcal{H}}$ vs. $m_V$ (right) for
$\tan\beta=0.5$ and $\tan\beta=4.0$ with $m_{\mathcal{H}}=70\;
\rm{GeV}$ and $m_{\mathcal{H}}=120\; \rm{GeV}$.} \label{fig:lambda}
\end{figure}

The $\mathcal{H}$-$D$-$D$ coupling $\lambda_{D\mathcal{H}}$ can be
derived from Eqs.~(\ref{eqn:anns}), (\ref{eqn:annf}) and
(\ref{eqn:annv}) for the scalar, fermion and vector DM,
respectively. They should satisfy the restricted region of
$\sigma_{ann} v_{rel}$ shown in Fig.~\ref{xf}. In
Fig.~\ref{fig:lambda} we show the $\lambda_{D\mathcal{H}}$ as a
function of $m_D$ for $\tan\beta=0.5$ and $\tan\beta=4.0$ with
different Higgs masses in the case of different dark matter spins.
They produce the correct central value of dark matter relic
abundance.

The scalar coupling $\lambda_{S\mathcal{H}}$ is displayed in the top panel of Fig.~\ref{fig:lambda}. From Eq.~(\ref{eqn:anns}), we know the scalar DM annihilation cross section approximately scales as
\begin{eqnarray}
\sigma_{ann}v_{rel}(S)\sim {3\lambda_{S\mathcal{H}}^2m_b^2\tan^2\beta\over 4\pi m_\mathcal{H}^4}
\end{eqnarray}
because the total decay width of Higgs is negligible compared with the Higgs mass in this case. Thus the coupling $\lambda_{S\mathcal{H}}$ is almost independent of the dark matter mass when Higgs mass is relatively large. With the same Higgs mass but different $\tan\beta$, the only factor that changes in the annihilation cross section is
$\Gamma(\tilde{\mathcal{H}}\rightarrow X_i)$ which is dominated by $\tilde{\mathcal{H}}\rightarrow b\bar{b}$.  Because the partial decay width of $\mathcal{H}\rightarrow b\bar{b}$ is proportional to $\tan^2\beta$, the smaller $\lambda_{S\mathcal{H}}$ is required to achieve the correct annihilation cross section for larger $\tan\beta$. On the other hand, the annihilation cross section is always suppressed by the Higgs mass. For the same $\tan\beta$, therefore, the coupling $\lambda_{S\mathcal{H}}$ should be smaller with decreasing Higgs mass.



For the fermionic dark matter, the dark matter annihilation is always P-wave suppressed.
Thus the annihilation cross section is about one order of magnitude larger than the S-wave case, which in turn requires a large $\lambda_{F\mathcal{H}}$.
Such a large $\lambda_{F\mathcal{H}}$ leads to a large Higgs decay total width dominated by
the invisible mode. The annihilation cross section is thus
\begin{equation}
\label{eqn:annfsim}
\sigma_{ann}v_{rel}(F)/v_{rel}^2\sim {3\lambda_{F\mathcal{H}}^2 m_F^2 m_b^2\tan^2\beta/(8\pi v_0^2)\over m_\mathcal{H}^4+m_\mathcal{H}^4\lambda_{F\mathcal{H}}^4/(16\pi)^2}.
\end{equation}
The annihilation cross section in the fermionic dark matter case
also has similar features with respect to DM mass, $\tan\beta$ and $m_{\mathcal{H}}$ as the scalar case. More importantly, at certain $\lambda_{F\mathcal{H}}$ and with some small $m_F$, small $\tan\beta$ and large $m_{\mathcal{H}}$, it reaches some value which does not necessarily fall into the restricted region as shown in bottom left panel of Fig.~\ref{fig:lambda}. We find that the annihilation cross section of fermion singlet DM into SM particles with any non-SM Higgs mass larger than 135 GeV would be
too small. As a result, the fermion singlet does not apply for the degenerate case with heavy $H^0$ in scenario I of THDM.

As seen from Eq.~(\ref{eqn:gammav}), the invisible decay of vector singlet DM is hugely enhanced by $m_{\mathcal{H}}^3/m_V^4$.
Its annihilation cross section is thus approximately
\begin{equation}
\label{eqn:annfsim}
\sigma_{ann}v_{rel}(V)\sim {2\lambda_{V\mathcal{H}}^2 m_b^2\tan^2\beta/(8\pi)\over m_\mathcal{H}^4+m_\mathcal{H}^8\lambda_{V\mathcal{H}}^4v_0^4/(128\pi m_V^4)^2}.
\end{equation}
Since this is pure S-wave annihilation, the coupling constant
$\lambda_{V\mathcal{H}}$ stays as large as in the scalar case. The
scaling behavior of the annihilation cross section with respect to
$\lambda_{V\mathcal{H}}$, however, is similar to the fermionic DM
case because of the large invisible decay width. This feature also
leads to the discontinued curves shown in the bottom right panel of
Fig.~\ref{fig:lambda}. The reachable limit of non-SM Higgs mass in
this case is also about 135 GeV.

%

\subsection{Coupling strength of the $D$-proton elastic scattering: $f_{Dp}$ }
\begin{figure}[htb]
\begin{center}
\includegraphics[scale=1,width=0.49\textwidth]{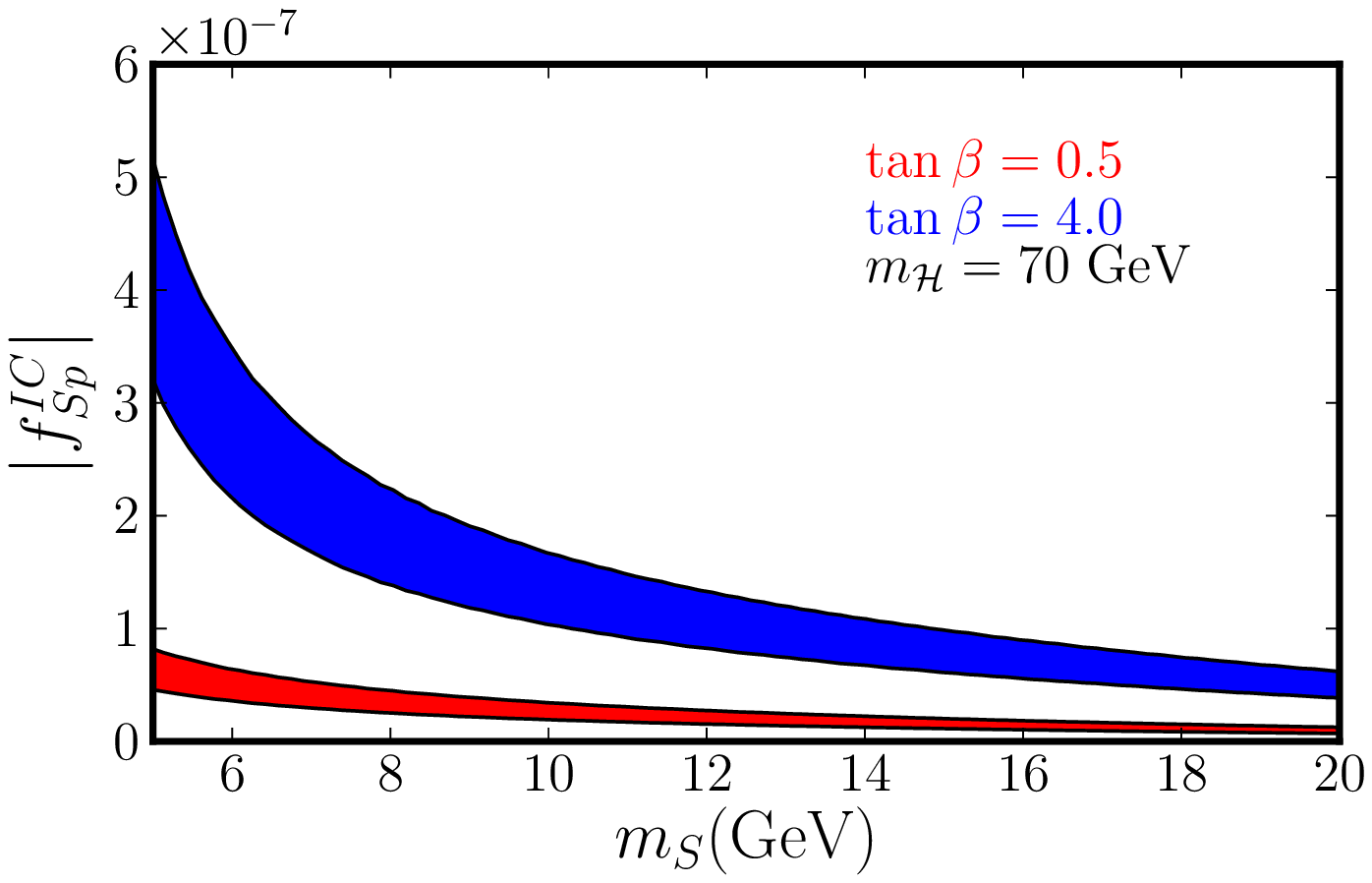}
\hfill
\includegraphics[scale=1,width=0.49\textwidth]{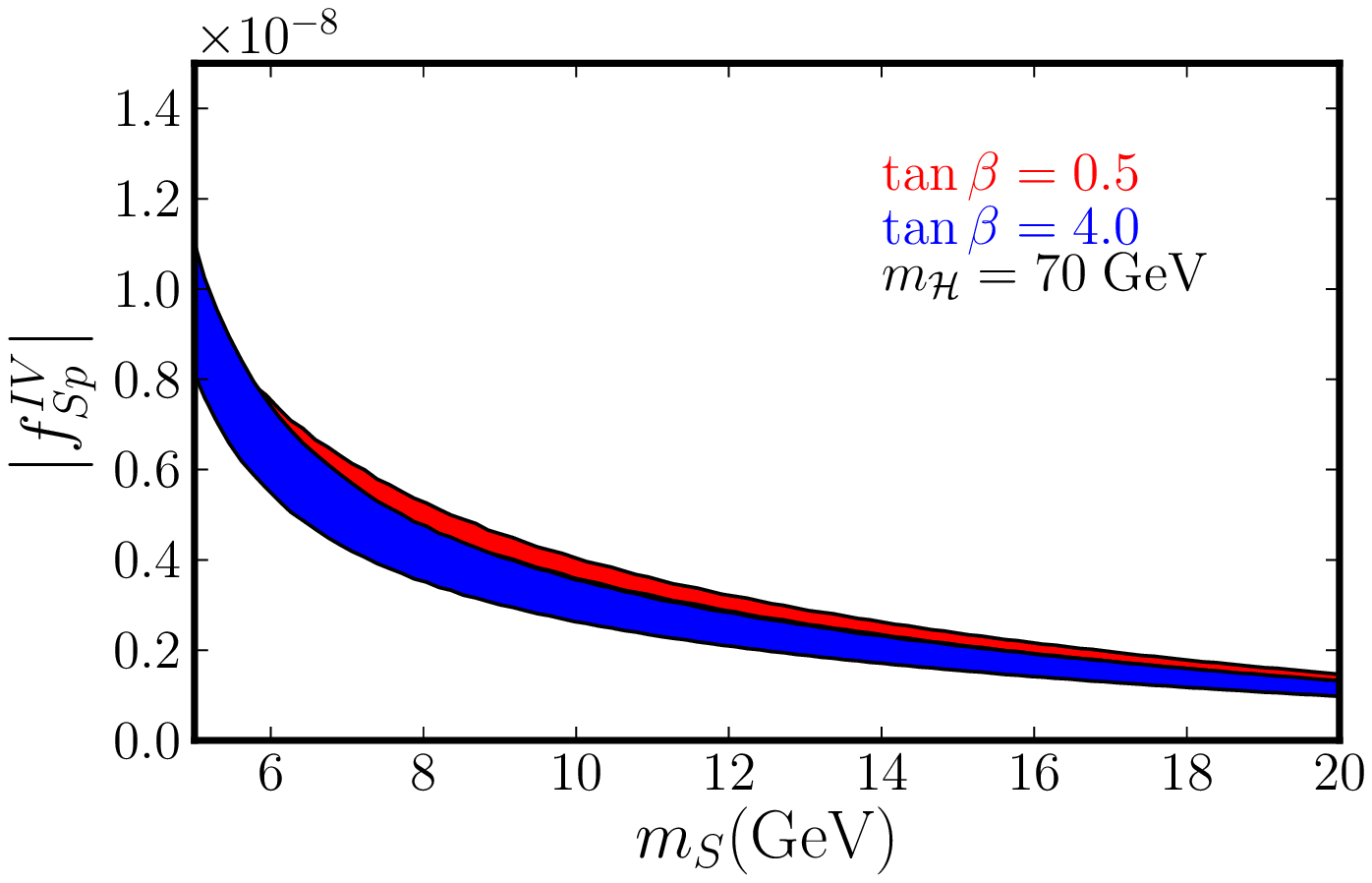}\\
\includegraphics[scale=1,width=0.49\textwidth]{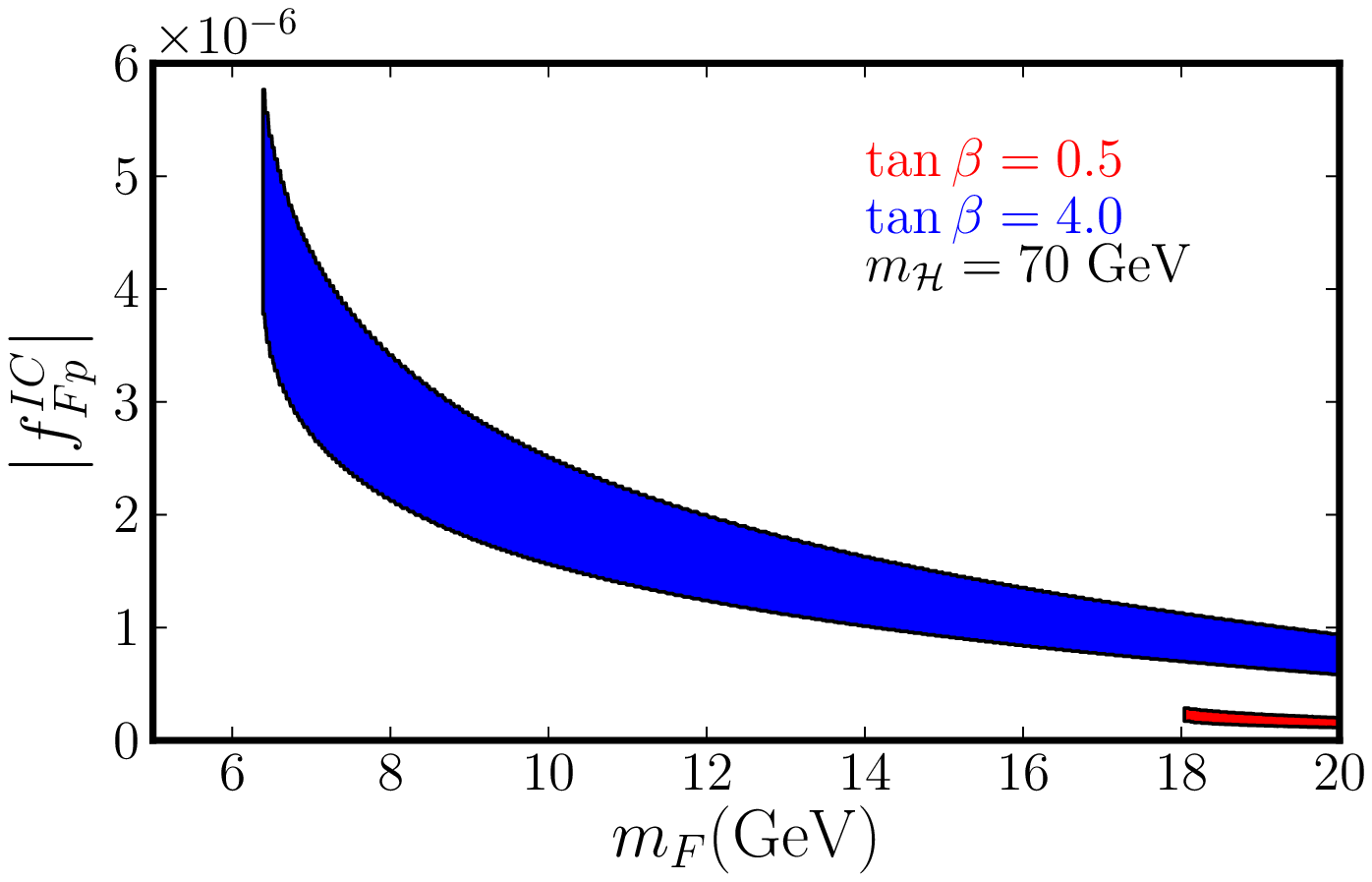}
\hfill
\includegraphics[scale=1,width=0.49\textwidth]{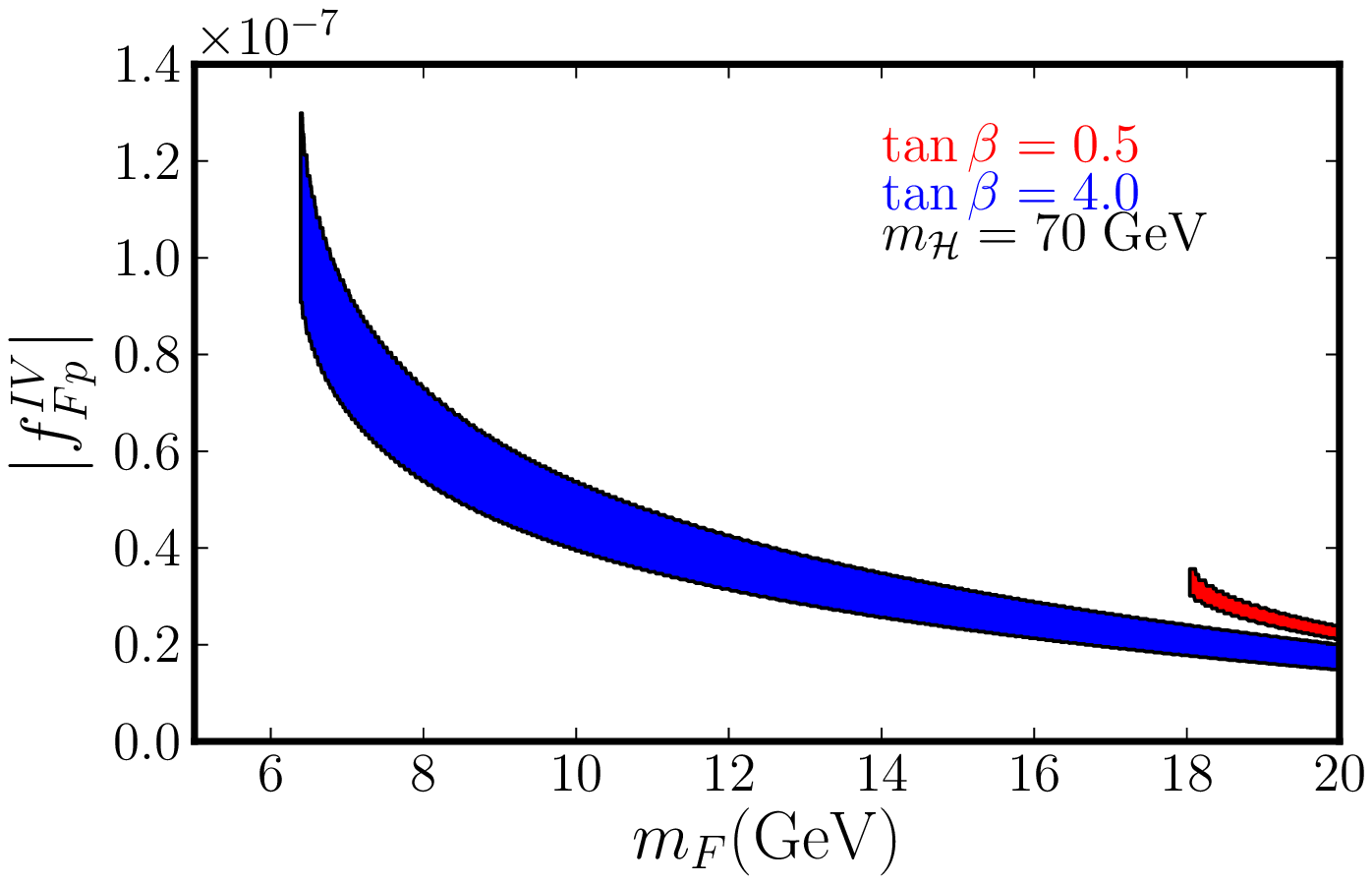}\\
\includegraphics[scale=1,width=0.49\textwidth]{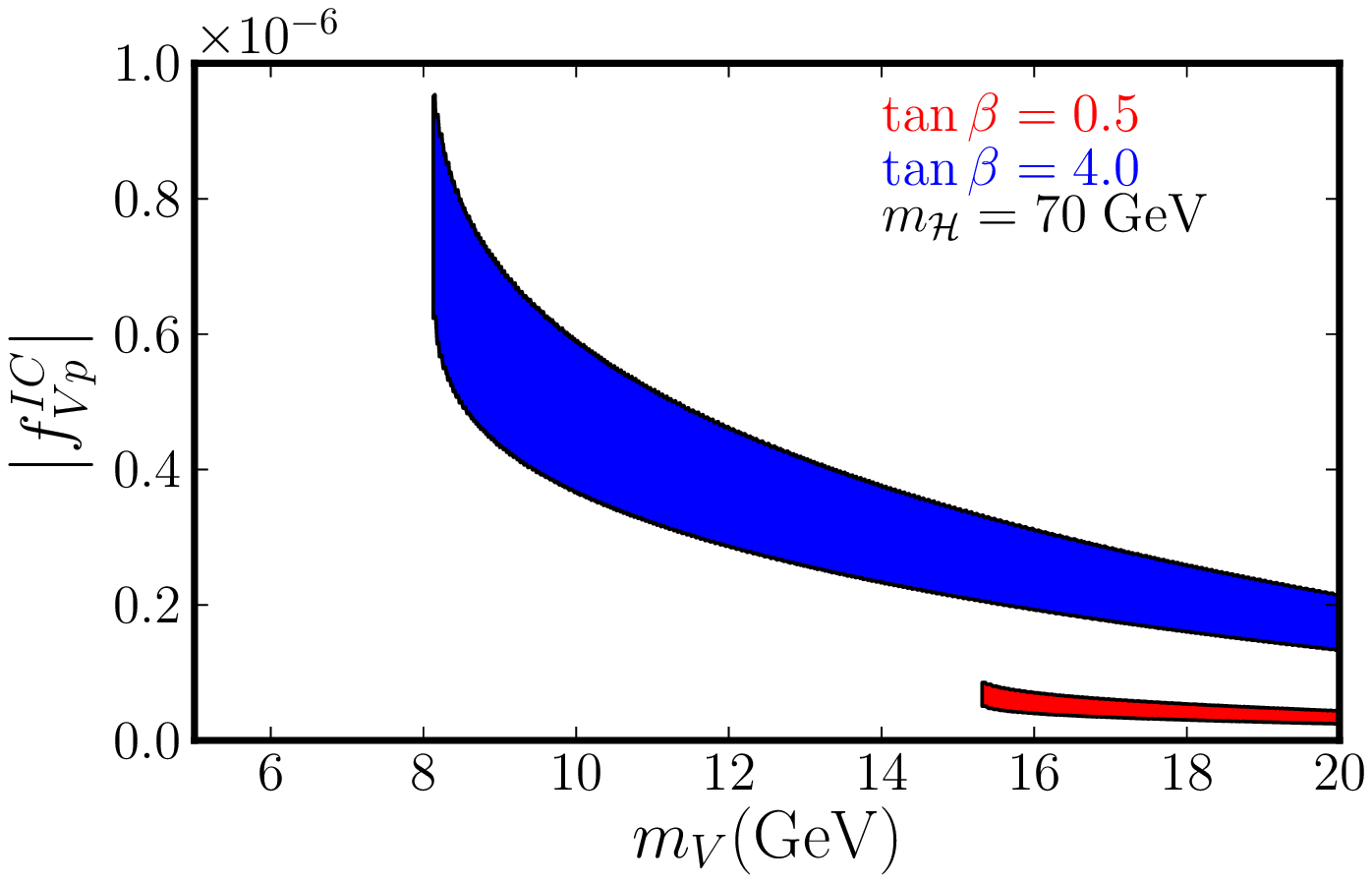}
\hfill
\includegraphics[scale=1,width=0.49\textwidth]{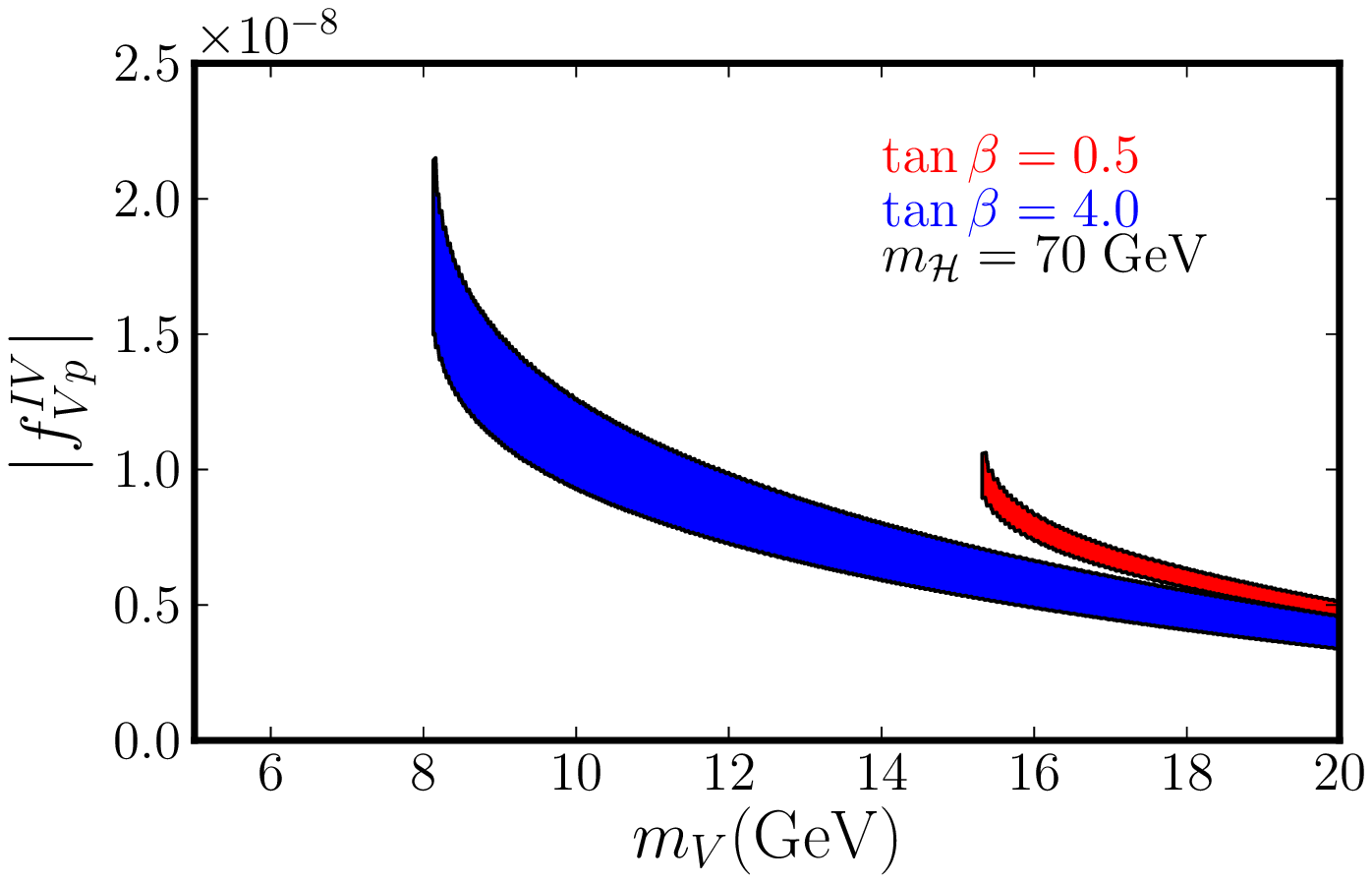}
\end{center}
\caption{Top: $f_{Sp}$ vs. $m_S$ for $\tan\beta=0.5$ and
$\tan\beta=4.0$ with $m_{\mathcal{H}}=70\;\rm{GeV}$. Middle:
$f_{Fp}$ vs. $m_F$ for $\tan\beta=0.5$ and $\tan\beta=4.0$ with
$m_{\mathcal{H}}=70\;\rm{GeV}$. Bottom: $f_{Vp}$ vs. $m_V$ for
$\tan\beta=0.5$ and $\tan\beta=4.0$ with
$m_{\mathcal{H}}=70\;\rm{GeV}$. Left column: IC case with
$f_{Dn}/f_{Dp}=1$; Right column: IV case with
$f_{Dn}/f_{Dp}=-0.64$.}
\label{fig:fDp}
\end{figure}

Combining the coupling $g_{pp\mathcal{H}}$ and $\lambda_{D\mathcal{H}}$ solved in previous subsections, we now proceed to study the elastic cross section of the singlet DM interacting with proton. The elastic cross section is determined by $f_{Dp}$ as a product of $g_{pp\mathcal{H}}$ and $\lambda_{D\mathcal{H}}$ as shown in Eqs.~(\ref{elS}), (\ref{elF}) and (\ref{elV}).
In Fig.~\ref{fig:fDp}, we display $f_{Dp}$ vs. $m_D$ for different $\tan\beta$ with and without isospin violation.
From Fig.~\ref{fig:gppH} and Fig.~\ref{fig:lambda}, we see that the $\lambda_{D\mathcal{H}}$
decreases with increasing $\tan\beta$ while the absolute value of $g_{pp\mathcal{H}}$ increases in both IC and IV cases.
In particular, $g_{pp\mathcal{H}}$ in the IC case increases more quickly than the decreasing of the $\lambda_{D\mathcal{H}}$ with increasing $\tan\beta$,
which results in an increasing $f_{Dp}$ with increasing $\tan\beta$. In the IV case, however, the increasing of $g_{pp\mathcal{H}}$ is slower and it
leads to an opposite trend with $\tan\beta$ for all DM singlets. The overall scale of $f_{Dp}$ in the IV case is much smaller than IC case since the coupling $g_{pp\mathcal{H}}$ in the IV case is two orders of magnitude smaller. Moreover, $f_{Sp}$ for different masses of Higgs remains the same as the Higgs mass dependence is cancelled in Eq.~(\ref{elS}). We thus show $f_{Dp}$ for only one particular Higgs mass in Fig.~\ref{fig:fDp}

\subsection{Invisible decay of singlet DM}
\begin{figure}[htp]
\begin{center}
\includegraphics[scale=1,width=0.45\textwidth]{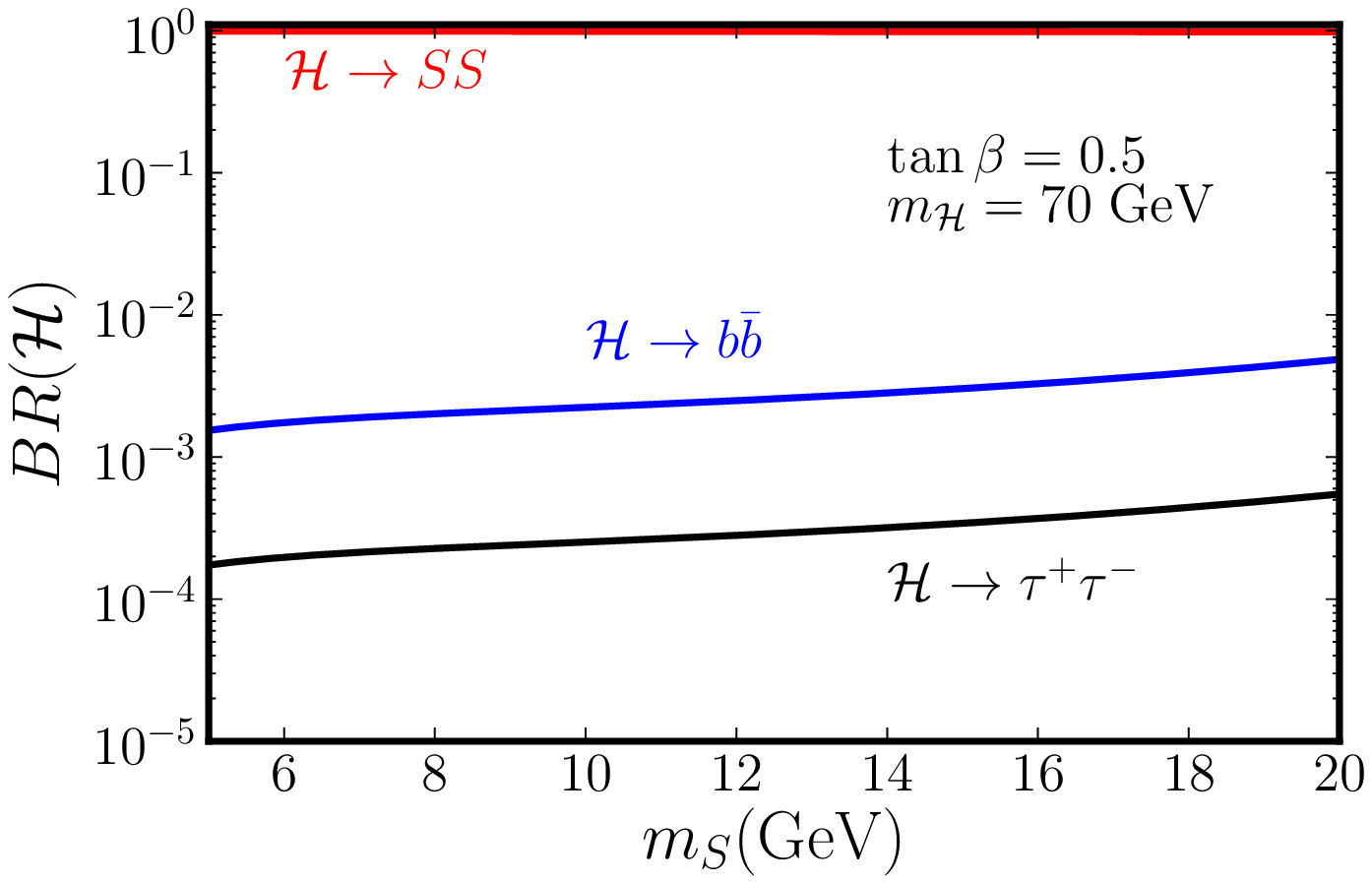}
\hfill
\includegraphics[scale=1,width=0.45\textwidth]{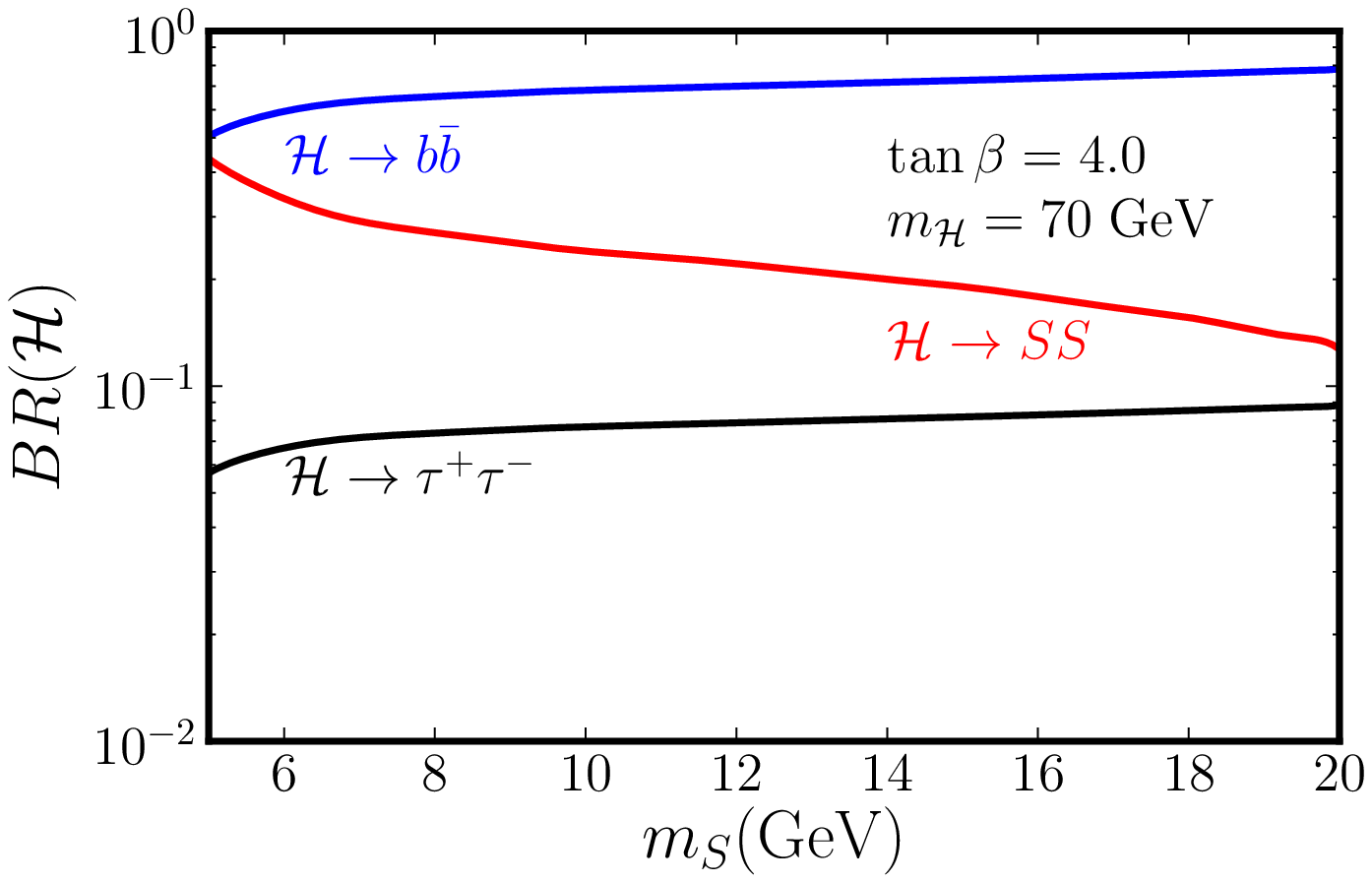}\\
\includegraphics[scale=1,width=0.45\textwidth]{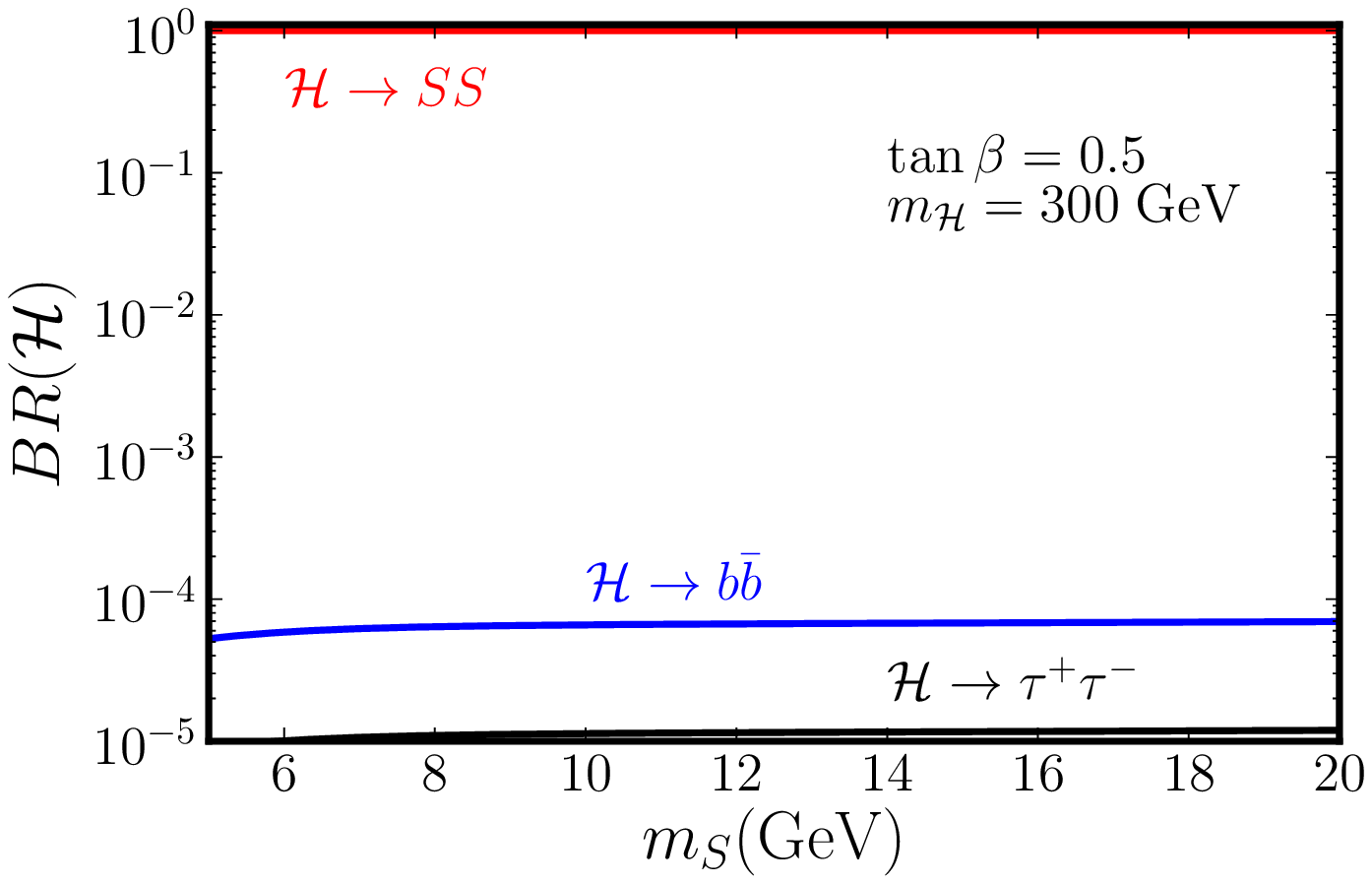}
\hfill
\includegraphics[scale=1,width=0.45\textwidth]{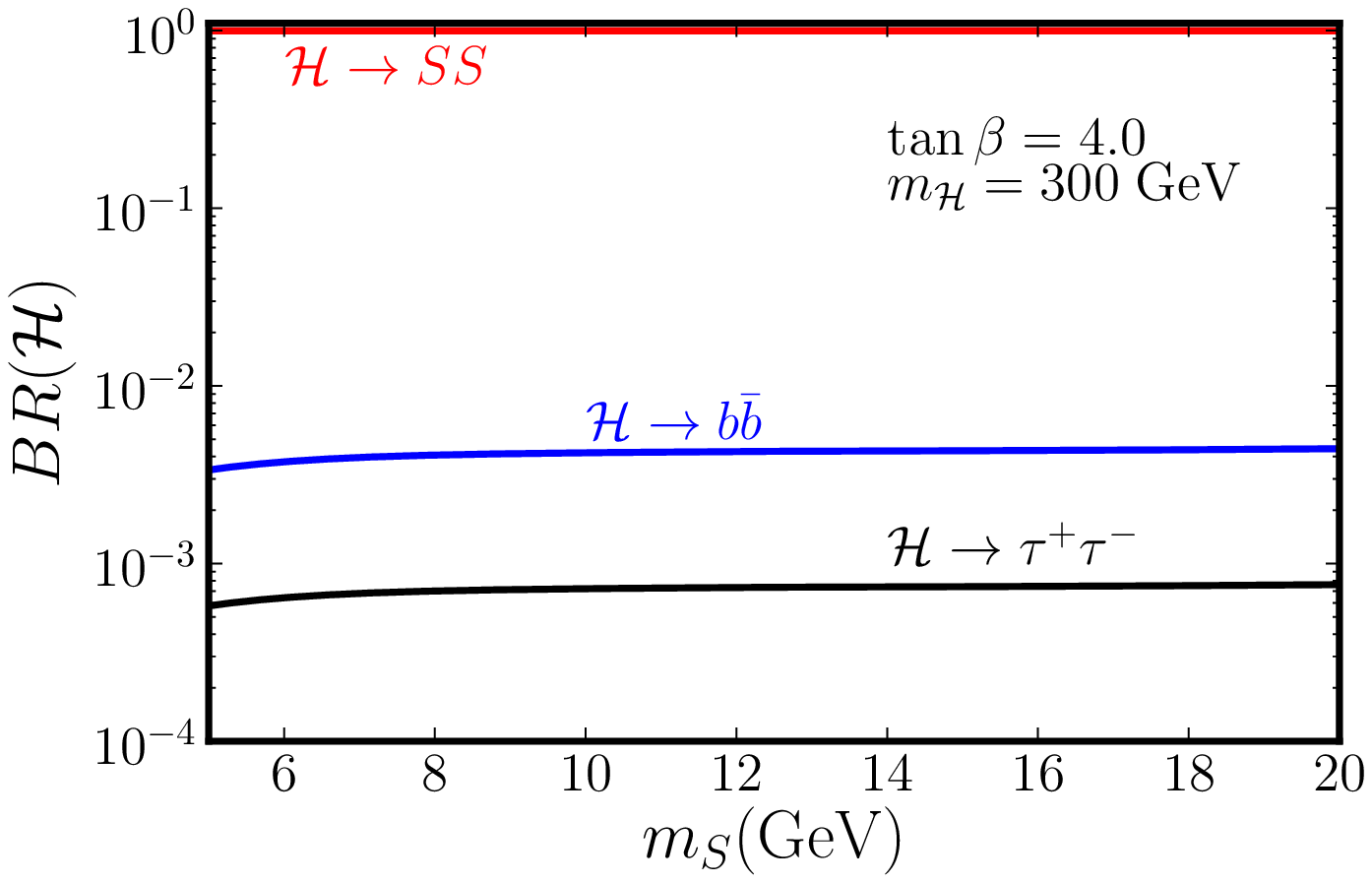}\\
\includegraphics[scale=1,width=0.45\textwidth]{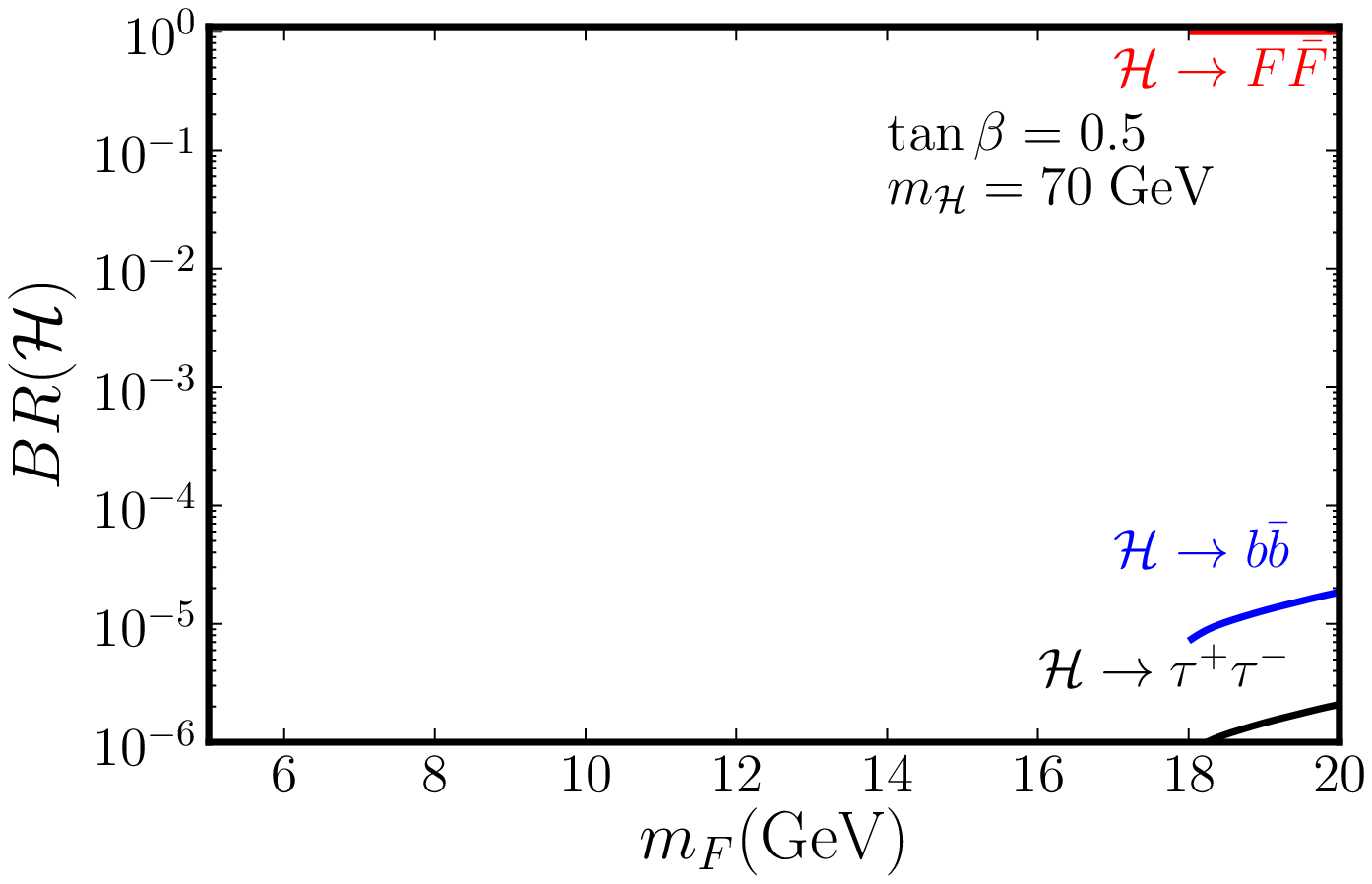}
\hfill
\includegraphics[scale=1,width=0.45\textwidth]{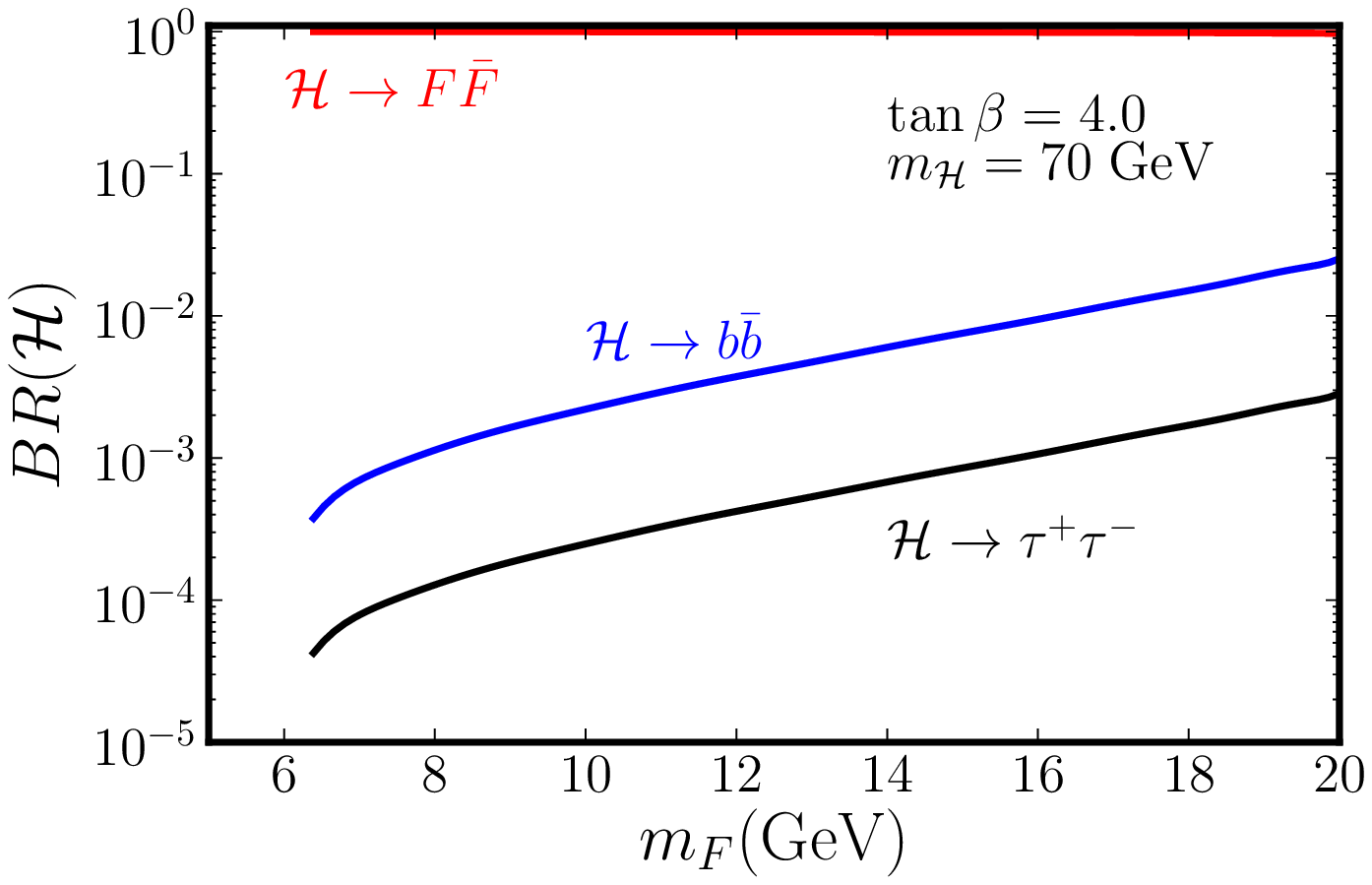}\\
\includegraphics[scale=1,width=0.45\textwidth]{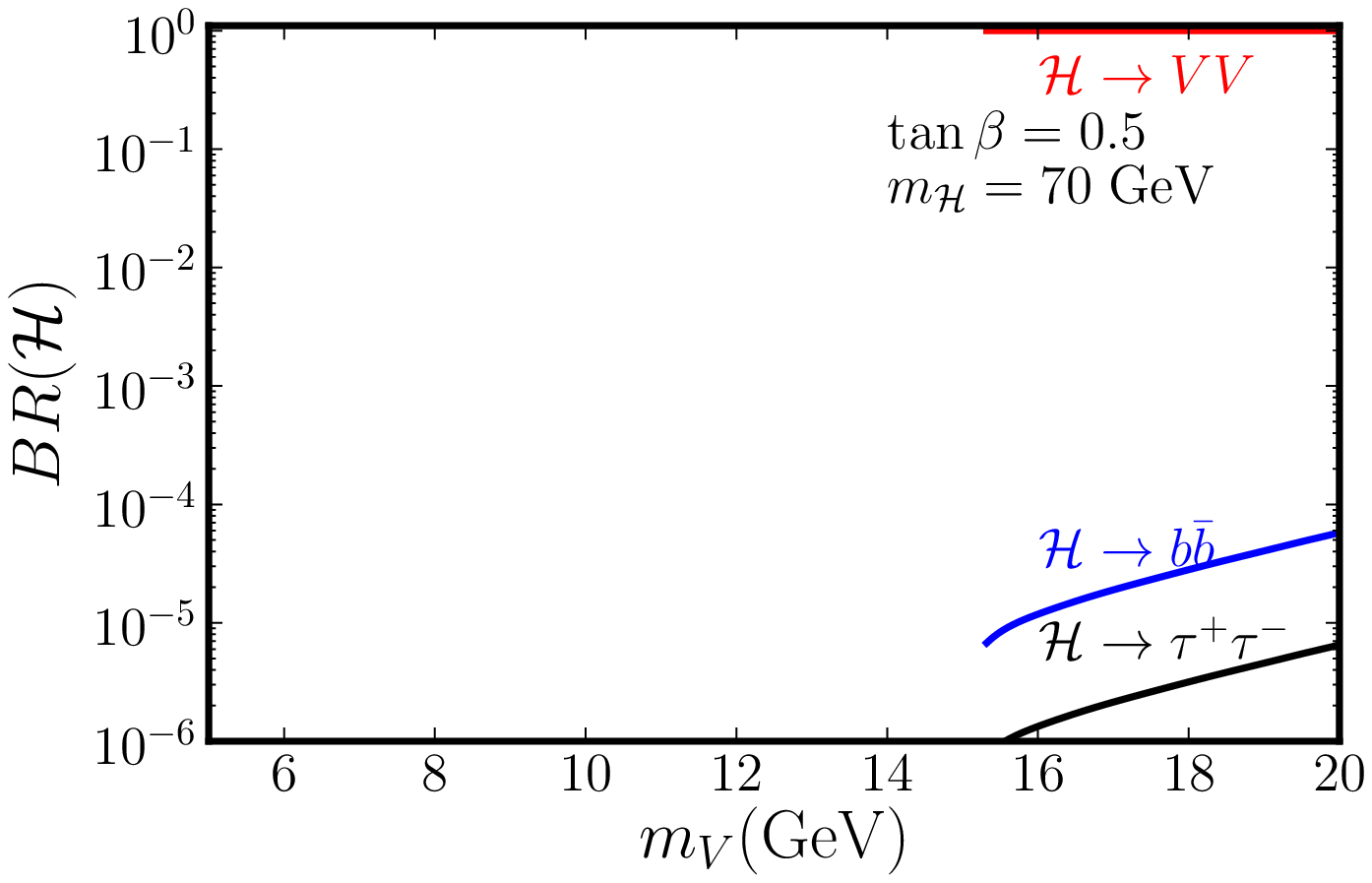}
\hfill
\includegraphics[scale=1,width=0.45\textwidth]{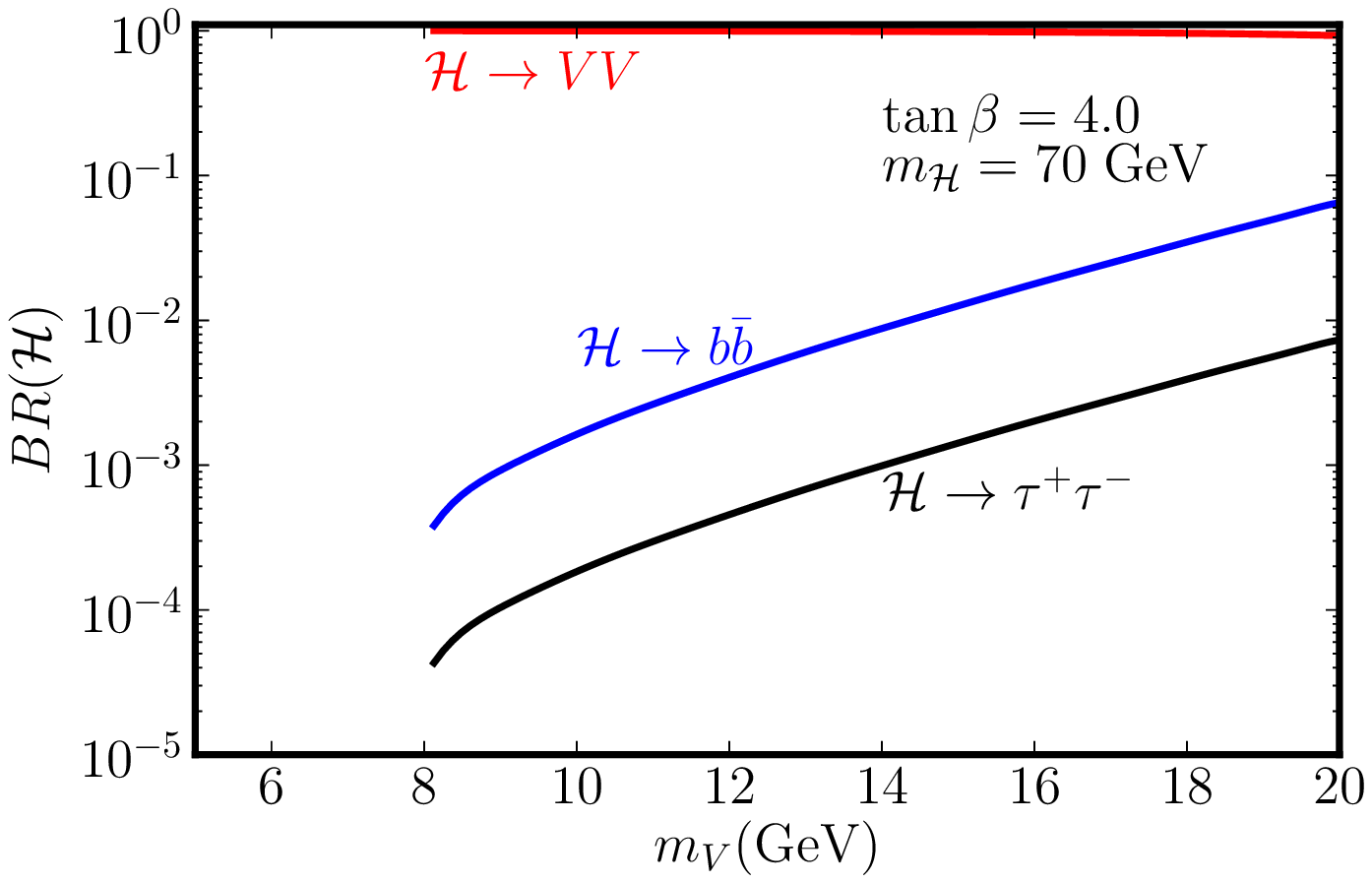}
\end{center}
\caption{$BR(\mathcal{H})$ vs. $m_S$ with $\tan\beta=0.5$ and
$\tan\beta=4$ for $m_{\mathcal{H}}=70\;\rm{GeV}$ (1st row) and
$300\;\rm{GeV}$(2nd row). $BR(\mathcal{H})$ vs. $m_F$ with
$\tan\beta=0.5$ and $\tan\beta=4$ for $m_{\mathcal{H}}=70\;\rm{GeV}$
(3rd row) and the same parameters for the vector case (4th row).}
\label{fig:higgsbr}
\end{figure}

Now we turn to the decay branching ratio of the Higgs $\mathcal{H}$,
as shown in Fig.~\ref{fig:higgsbr} for different Higgs mass,
$\tan\beta$ and singlet DM candidates. The branching ratios here are
calculated with the coupling $\lambda_{D\mathcal{H}}$ from
Fig.~\ref{fig:lambda}. The common feature is that the Higgs decay
width is significantly dominated by the invisible channel
$\mathcal{H}\rightarrow DD$. The other major contribution comes from
$\mathcal{H}\rightarrow b\bar{b}$ and $\mathcal{H} \rightarrow
\tau^+ \tau^-$. Such significant invisible decay in THDM can be
tested through mono-b jet process $gb\to b\mathcal{H}$.

\subsection{$D$-proton elastic scattering cross section $\sigma_{el}^p$}
\begin{figure}[htp]
\begin{center}
\subfloat[\label{dd:a}]{
\includegraphics[scale=1,width=0.49\textwidth]{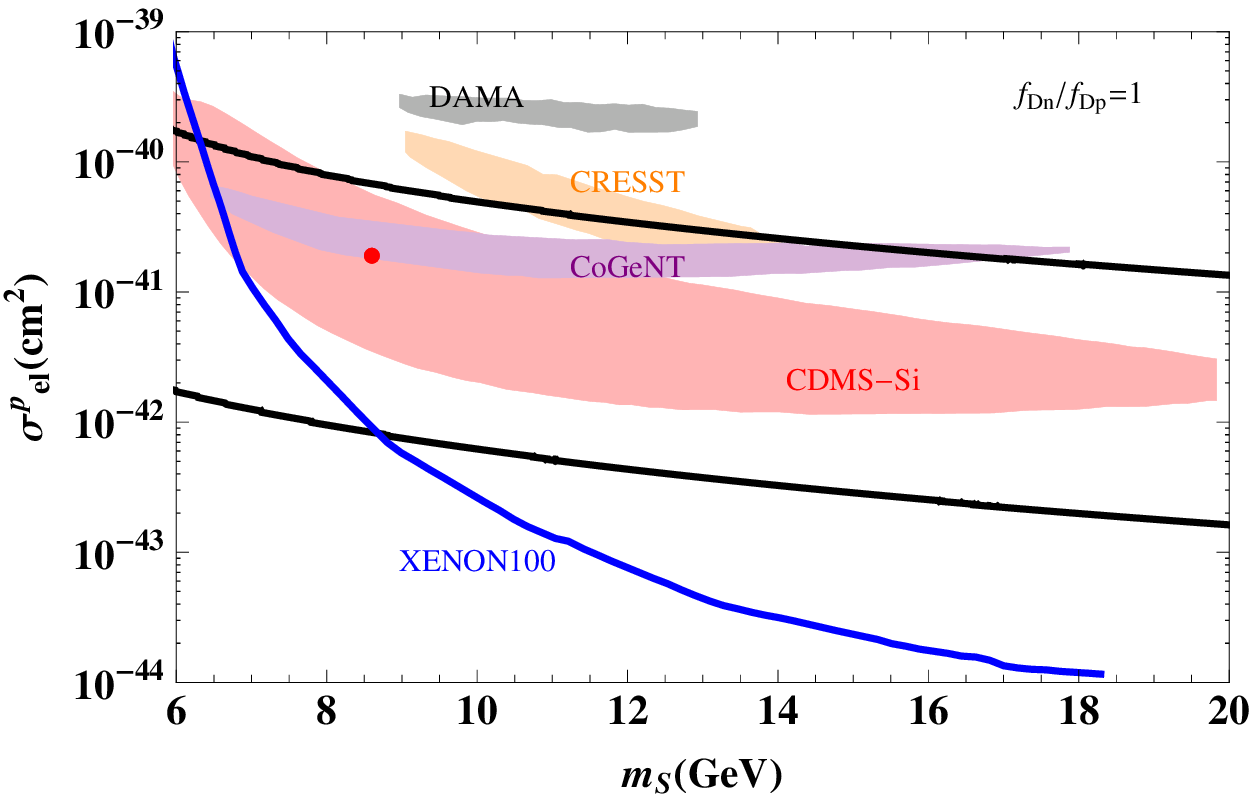}}
\hfill
\subfloat[\label{dd:b}]{
\includegraphics[scale=1,width=0.49\textwidth]{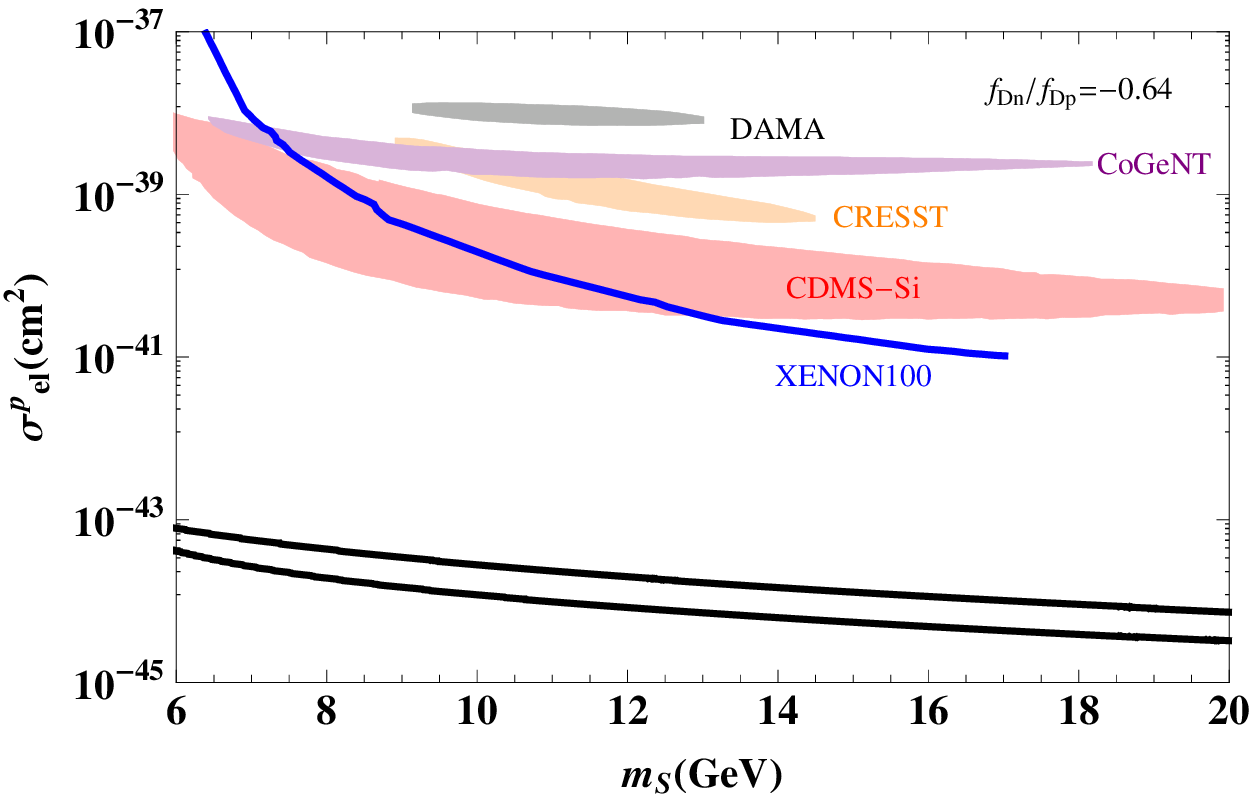}}\\
\subfloat[\label{dd:c}]{
\includegraphics[scale=1,width=0.49\textwidth]{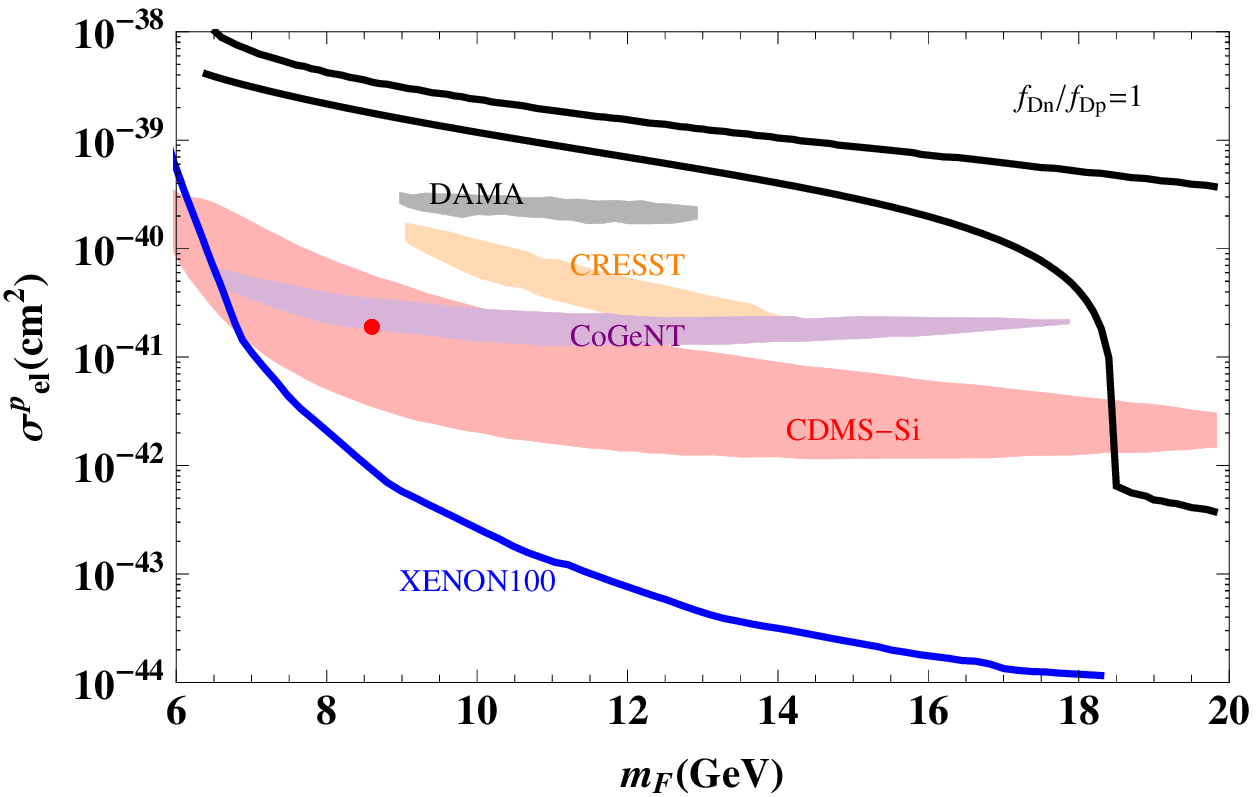}}
\hfill
\subfloat[\label{dd:d}]{
\includegraphics[scale=1,width=0.49\textwidth]{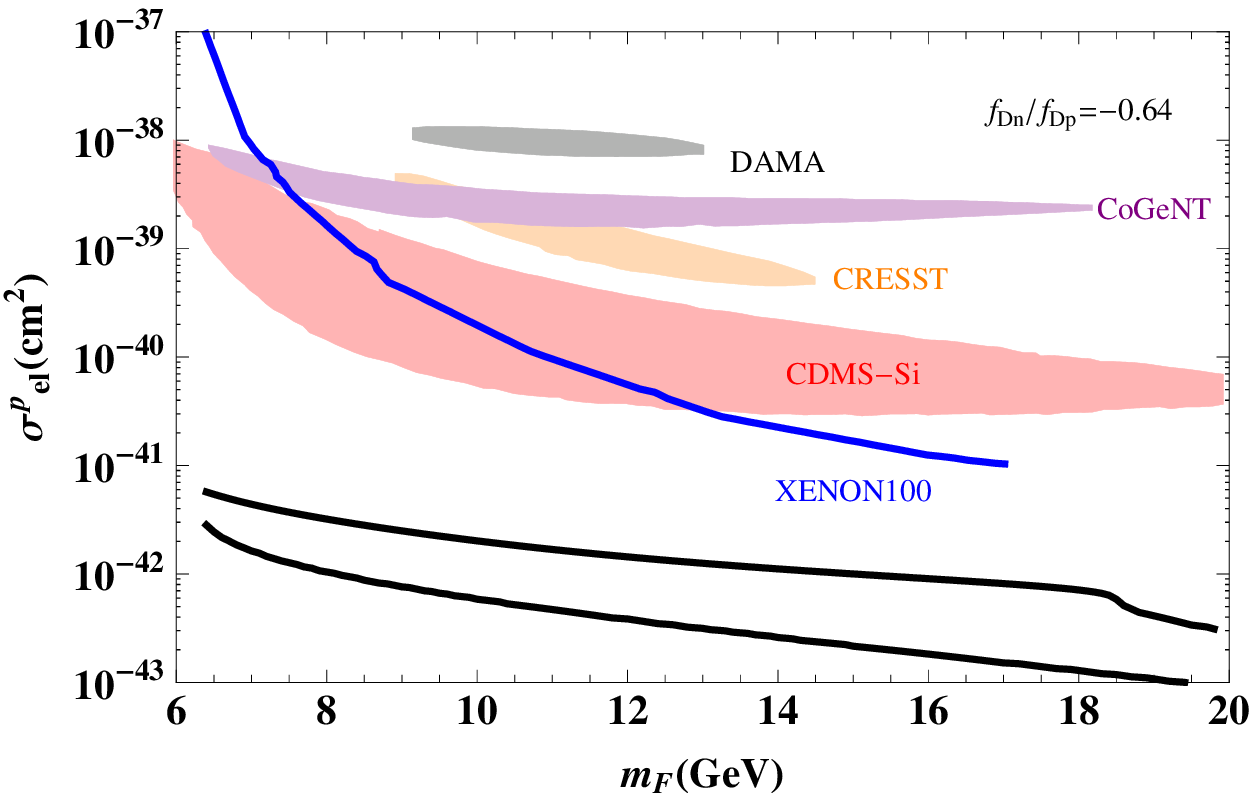}}\\
\subfloat[\label{dd:e}]{
\includegraphics[scale=1,width=0.49\textwidth]{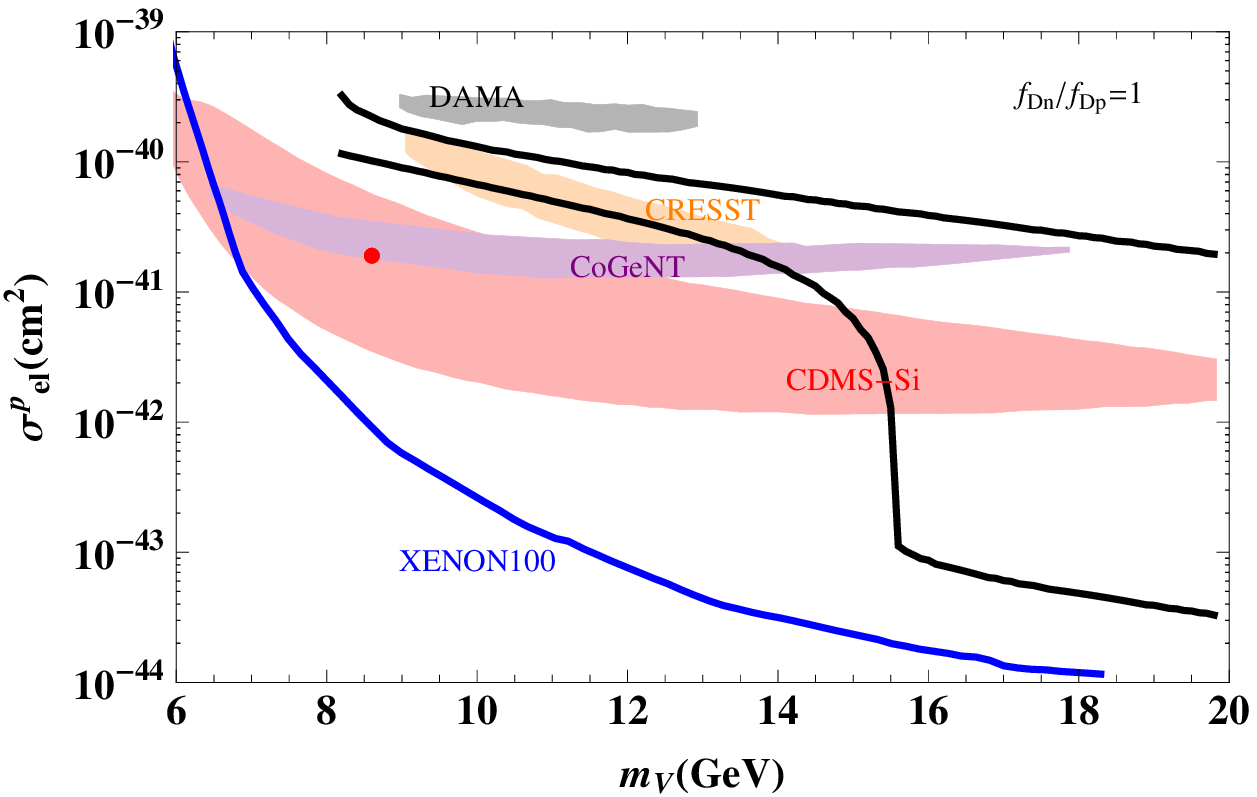}}
\hfill
\subfloat[\label{dd:f}]{
\includegraphics[scale=1,width=0.49\textwidth]{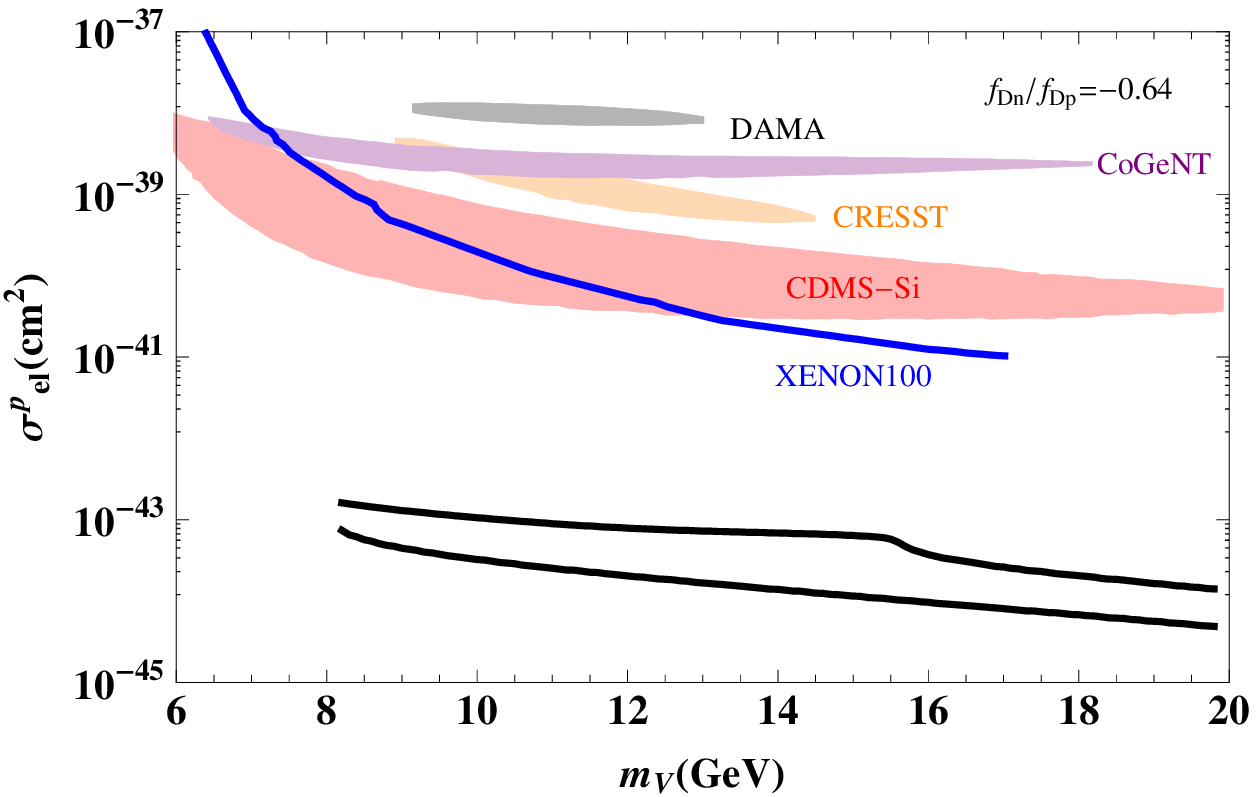}}
\caption{Top: The elastic cross section of scalar DM with proton.
Middle: The elastic cross section of fermionic DM with proton.
Bottom: The elastic cross section of vector DM with proton. Left
column: IC case with $f_{Dn}/f_{Dp}=1$; Right column: IV case with
$f_{Dn}/f_{Dp}=-0.64$. $m_\mathcal{H}$ is fixed to be 70 GeV and $\tan\beta$ varies from 0.5 to 4. The $90\%$ CL ROIs for CDMS-Si, CoGeNT, and
the $3(2)\sigma$ ROIs for DAMA (CRESST-II) are also plotted together
with the exclusion limits from XENON100. The red point denotes the best fitted result from CDMS-Si.}
\label{fig:DD}
\end{center}
\end{figure}

At last we gather all the necessary ingredients to calculate the elastic cross section of dark matter scattering with proton according to Eqs.~(\ref{elS}), (\ref{elF}) and (\ref{elV}). The uncertainties come from the measurement of the dark matter relic abundance $\Omega h^2$
and the coupling $g_{pp\mathcal{H}}$. The results for IC and IV case with various singlet DM candidates and $m_\mathcal{H}=70$ GeV are shown in Fig.~\ref{fig:DD}. For scalar singlet, the lower (upper) and upper (lower) limit of the elastic cross section is obtained with $\tan\beta=0.5$ and $\tan\beta=4.0$ in the IC (IV) case. As stated before, these limits apply for any values of $m_\mathcal{H}$ in this case.
Fig.~\ref{dd:a} shows that the restricted region of the parameter
space is consistent with most of the region of interests (ROI) of
CDMS-Si for scalar DM with IC, while the IV case (Fig.~\ref{dd:b})
has a two orders of magnitude smaller elastic cross section due to
the much small $g_{pp\mathcal{H}}$.

For fermionic singlet DM (see Figs.~\ref{dd:c} and \ref{dd:d}), the elastic cross section in the restricted region is about two orders of magnitude larger than the
scalar case, since the annihilation cross section is P-wave suppressed and a much larger $\mathcal{H}$-$D$-$D$ coupling is needed.
Similar to the scalar DM, the fermion elastic cross section increase with increasing $\tan\beta$ in the IC case and decreases in the IV case. For larger $m_\mathcal{H}$, the allowed region of elastic cross section is more restricted and only covers relatively large values of $m_F$.

In the vector DM case (see Figs.~\ref{dd:e} and \ref{dd:f}), the
elastic cross section is of the same order of magnitude as the
scalar DM case. But the restricted region is narrower and overlaps
much less with the ROIs of CDMS-Si than the scalar DM case. It is
because the enhanced dark matter decay width suppresses the total
annihilation cross section in some parameter space and the
restricted region is further constrained as already explained in
Sec.~\ref{subsec:lambdadh}.

\section{Conclusions}
\label{sec:conclusion}
The light WIMP dark matter is of special interests given the recent
events from direct detection experiments. We have studied the simple
extension of the minimal SM with two Higgs doublets and a SM gauge
singlet stabilized by $Z_2$ symmetry in the low mass region. We
focused on the Type II THDM with a vanishing coupling between the
singlet dark matter and the SM-like Higgs such that single DM can
only couple to the non-SM Higgs in pairs. The SM-like Higgs invisible decay is thus
consistent with LHC search. In this limit, the
only parameters of the model include the dark matter mass $m_D$, the
non-SM CP-even Higgs mass $m_\mathcal{H}$, $\tan\beta$ and the
$\mathcal{H}-D-D$ coupling $\lambda_{D\mathcal{H}}$.

To obtain appropriate dark matter relic abundance, we found in the
case of fermionic singlet, a large coupling $\lambda_{F\mathcal{H}}$
is required because of P-wave suppression. In the vector singlet
case, the coupling $\lambda_{V\mathcal{H}}$ remains at the same
order as that of the scalar case, but one order of magnitude smaller
than the fermion case.
The Higgs decay width is dominated by the invisible mode. In both
fermionic and vector DM case the large invisible decay width leads
to a suppressed annihilation cross section. In some parameter space,
the dark matter would overclose the universe.

We also considered the hadronic uncertainties in both isospin
conversing and isospin violating cases (especially the nearly
xenon-phobic case with $f_{Dn}/f_{Dp}=-0.64$). It turns out in this
model, with LHC favored parameter region, the Higgs-proton coupling
$g_{pp\mathcal{H}}$ in the IV case is almost two orders of magnitude
smaller than that in IC case. This feature would definitely suppress
the elastic scattering cross section between dark matter and
nucleon.

After scanning the allowed parameter space, we found that only the
scalar singlet without IV effects has a major overlap with the
region of interests of most direct detection experiments although it
is still in tension with XENON100. The elastic cross section in the
IC fermionic case is much larger than all the experimental ROIs due
to the large coupling $\lambda_{F\mathcal{H}}$ which is required to
compensate for the P-wave suppression. The vector case has less overlap
with the experimental ROIs because in some parameter space the large
invisible decay width leads to an overclosed universe. All the
scenarios in the IV case have small elastic cross section because of
the small Higgs-proton coupling.

\subsection*{Acknowledgment}
We would like to thank Robert Foot, Michael A. Schmidt, Shufang Su and Felix Yu for useful discussions.
T.L. would also like to thank Fermilab for the hospitality during which part of this work was carried out.
This work was supported in part by the Australian Research Council.

\appendix

\section{Higgs-nucleon couplings}
\label{app:gppH}

For isospin-conserving case ($f_{Dn}/f_{Dp}=1$), the $g_q^N$s ($N=n,p$) are~\cite{hedarkon,ic1}
\begin{eqnarray}
g_u^N&=&g_d^N={\sigma_{\pi N}\over 2v_0}, \ g_s^N={m_N-m_B-\sigma_{\pi N}\over v_0}, \ g_{Q}^N={2m_B\over 27 v_0},\\
m_B&=&-\sigma_{\pi N}{2m_K^2+m_\pi^2\over
2m_\pi^2}+{(m_\Xi+m_\Sigma)(2m_K^2-m_\pi^2)-2m_Nm_\pi^2\over
4(m_K^2-m_\pi^2)}.
\end{eqnarray}

For isospin-violating case ($f_{Dn}/f_{Dp}\neq 1$), they are~\cite{update,ic1,pdg,lattice1,lattice2}
\begin{eqnarray}
g_q^N&=&{B_q^Nm_q\over v_0},  \ q=u,d,s;  \ \ \ g_{Q}^N={2\over 27 v_0}\left(m_N-\sum_q m_qB_q^N\right), \ Q=c,b,t,\\
B_u^N&=&{2\sigma_{\pi N}\over m_u(1+{m_d\over m_u})(1+{B_d^N\over B_u^N})}, \ B_d^N={2\sigma_{\pi N}\over m_d(1+{m_u\over m_d})(1+{B_u^N\over B_d^N})}, \ B_s^N={{m_s\over m_d}\sigma_{\pi N}(1-{\sigma_0\over \sigma_{\pi N}})\over m_s(1+{m_u\over m_d})},\\
{B_d^p\over B_u^p}&=&{2+(z-1)(1-{\sigma_0\over \sigma_{\pi N}})\over 2z-(z-1)(1-{\sigma_0\over \sigma_{\pi N}})},  \quad \ B_u^n=B_d^p, \quad  \ B_d^n=B_u^p,\\
{m_u\over m_d}&=&0.38-0.58, \quad \ {m_s\over m_d}=17-22, \quad \ z \equiv{B_u^p-B_s^p\over B_d^p-B_s^p}=1.49, \\
\sigma_0&=& 58\pm 9 \ {\rm MeV}, \quad  \sigma_{\pi N}=58\pm 9 \ {\rm MeV}.
\end{eqnarray}


\end{document}